\documentclass[intlimits,twoside,a4paper]{article}

\usepackage{amsmath,amssymb}
\usepackage{graphicx}

\usepackage{subfigure}
\usepackage[T2A]{fontenc}
\usepackage[cp1251]{inputenc}

\usepackage[eqsecnum]{cmpj2}

\issue{2015}{18}{4}{43702}
\doinumber{10.5488/CMP.18.43702}

\title[Phase transitions in BFH model]%
{Phase transitions in Bose-Fermi-Hubbard model in the heavy fermion limit: Hard-core boson approach}
\author[I.V. Stasyuk, V.O. Krasnov]{I.V. Stasyuk, V.O. Krasnov}
\address{
Institute for Condensed Matter Physics of the National Academy of Sciences of Ukraine,\\
1 Svientsitskii St., 79011 Lviv, Ukraine}

\date{Received July 2, 2015}
\authorcopyright{I.V. Stasyuk, V.O. Krasnov, 2015}

\begin{document}

\maketitle

\begin{abstract}
Phase transitions are investigated in the Bose-Fermi-Hubbard model in the mean field and hard-core boson approximations for the case
of infinitely small fermion transfer and repulsive on-site boson-fermion interaction.
The behavior of the Bose-Einstein condensate order parameter and grand canonical potential  is analyzed as functions of the chemical potential of bosons at zero temperature.
The possibility of change of order of the phase transition to the superfluid phase in the regime of fixed values of the chemical potentials of Bose- and
Fermi-particles is established. The relevant phase diagrams are built.
\pacs 71.10 Fd, 71.38
\end{abstract}

\section{Introduction}
\label{sec:1}

Physics of many-particle systems with strong short-range correlations is one of the important research fields.
A new outbreak of activity in this area is associated with manifestations of peculiar properties of ultracold atomic systems in traps at
imposition of periodic field formed by the interference of coherent laser beams. In the optical lattices formed this way, there occurs a transition to the state with Bose-Einstein (BE) condensate at very low temperatures $(T<10^{-7}-10^{-8}~\textrm{K})$ in the case of Bose-atoms. Experimentally, this effect was originally observed
by Greiner and others in 2002--2003 \cite{gr1,gr2} in the system of $^{87}$Rb atoms. Bose-Einstein condensation occurs in this case by the phase
transition of the 2nd order from the so-called Mott insulator phase (MI-phase) to the superfluid phase (SF-phase). The theoretical description of the phenomenon
is based on the Bose-Hubbard model \cite{fish,jak}, which takes into account two main factors that determine the thermodynamics and energy spectrum of the system of Bose-particles~--- tunneling between the neighbouring minima of potential in the lattice and short-range on-site repulsive interaction of Hubbard type.
A lot of investigations \cite{vOst,elst,batr,clark,dirk,seng,oha,hub,kona,ista1}
are dedicated to the construction of phase diagrams that define the conditions of the existence of
SF-phase at $T=0$ and at non-zero temperatures as well as to the study of different aspects of one-particle spectrum (see, also, reviews \cite{bloch1,bloch2,lew1}).

Along with ultracold atomic Bose-systems, the boson-fermion mixtures in optical atomic lattices are also actively researched. Their experimental implementation
(e.g., spin-polarized mixtures of $^{87}$Rb--$^{40}$K atoms \cite{best,gunt,osp1}) made it possible to see the MI-SF transition in the presence of Fermi-atoms.
An important factor that has been observed is the fading of the coherence of bosons and the decay of condensate fraction in SF-phase in a certain range of the values of thermodynamic parameter
(chemical potential of bosons or temperature) influenced by the interaction with fermions. This is reflected in the change of conditions of existence of SF- and MI-phases and the shift of curve of SF-MI transition on the phase diagram. Renormalization (due to self-trapping) of
tunneling hopping of bosons \cite{best,osp1,luh}, influence of higher boson bands \cite{tewar,mering}, the intersite interactions
(including the so-called correlated hopping) \cite{jur} are considered to be possible explanations of this effect. An important feature of boson-fermion mixtures is the possible existence of new quantum phases such as
CDW-type phase (with the particle density modulation), supersolid (SS) phase (with the spatial modulation of density as well as the order parameter of condensate),
SF$_\textrm{f}$ phase (with the condensate of fermion pairs), and their various combinations (see, in particular, \cite{buk}). Another interesting factor that should be taken into consideration is the formation of the so-called fermion composites \cite{fehr1,cramer}, which are caused by the fermion pairing with one or more bosons
(or one or more boson holes) due to their effective attraction (or repulsion). Interaction of interspecies in this case can be changed \cite{fesh} using the Feshbach resonance \cite{inou}.
An asymmetry between the attraction and repulsion cases in the behavior of boson-fermion mixtures and in the phase diagrams \cite{best,alb1} is observed here.

The Bose-Fermi-Hubbard model is used to describe the boson-fermion mixtures in optical lattices. This model and its microscopic justification were proposed in \cite{alb1}.
Following this, in \cite{fehr2,lew2} the phase diagrams at $T=0$ (ground state diagrams) in the mean-field approximation were constructed. Moreover,
the atom transfer was taken into account using the perturbation theory; the effective static interaction between bosons was included
(in the cases $J_\textrm{B}=J_\textrm{F}=0$; $J_\textrm{B}\neq 0$, $J_\textrm{F}=0$; $J_\textrm{B}=J_\textrm{F}=J$, where $J_\textrm{B}$, $J_\textrm{F}$ are parameters of the atom transfer). The areas of the existence of phases with composite fermions that contain a different number of bosons (or boson holes) were found. The analysis was performed in the regime of fixed fermion density.

In \cite{buh2} as well as in subsequent studies \cite{polak,refa}, it was shown within the BFH model that the direct boson-fermion interaction can lead
to an effective dynamic interaction between bosons through fermionic field. This gives the appearance of instability; when $\vec{q}=0$~--- in regard to phase separation, and when $\vec{q}=\vec{k}_\textrm{DW}$~--- in regard to spatial modulation and SS phase formation (which in the case of half-filling for
fermions and the increase of energy of their repulsion off bosons becomes a CDW phase \cite{tit1}).
The mechanisms of SS phase arising were studied in some other works (see, in particular \cite{buk}). Bose-Einstein condensate enhances the s-wave pairing
of fermions, while uncondensed bosons  contribute to the appearance of CDW. At half-filling for fermions, SF$_\textrm{f}$ -phase competes with antiferromagnetic ordering \cite{anders}.

On the other hand, at the spin degeneracy, the reverse effect is possible when the pairing of fermions is induced by bosons. This situation is analogous to the formation of Cooper pairs in the BCS model, as was shown in several papers (see, \cite{bijls,heisel,viver}).
It results in the creation of a  SF$_\textrm{f}$ phase, where the role of the superfluid component belongs to fermion pairs; the corresponding phase diagrams are constructed in \cite{buk}.
Integration over fermionic variables also provides an additional static interaction  $U_{BB}$ between bosons, which promotes MI~$\rightarrow$~SF transition
or suppresses it.
To a large extent it depends on the mass ratio of Bose- and Fermi-atoms (i.e., on the ratio of the hopping parameters $t_\textrm{F}/t_\textrm{B}$).
The phase diagram obtained by functional integration and the Gutzwiller approach for the cases of ``light'' (a mixture of $^{87}$Rb--$^{40}$K atoms) and ``heavy''
(a mixture of $^{23}$Na--$^{40}$K atoms) fermions is presented in \cite{polak}.

Other aspects of the fermion influence on the MI~$\rightarrow$~SF  phase transition in a boson subsystem were investigated in
\cite{tewar,mering,jur,buk,fehr1,cramer,fesh,inou,alb1,fehr2,lew2,buh2,polak,refa,tit1,anders,bijls,heisel,viver,lutch1}. It is shown, in particular, that virtual
transitions of bosons under the effect of interaction $U_\textrm{BF}$ through their excited states in the optical lattice potential minima lead to an extension of the MI phase region at $T = 0$ (the shift of the curve of the phase transition in the ($t$, $\mu$) plane towards larger values of the $t/U$ ratio). A similar role is played
by retardation at the boson screening by fermions and ``polaron effect'' \cite{math}. This is manifested by the reduction of the parameter of the transfer of bosons $t_\textrm{B}$ and by their slowing down.
However, for heavy fermions, the SF phase region broadens for the case of intermediate temperatures \cite{refa}. Interparticle interactions, in particular the
so-called bond-charge interaction, can have a significant effect on the transfer of bosons as shown in \cite{jur}. This also can lead to the shift of
transition from MI to SF phase.

A separate direction of theoretical description of boson atoms and boson-fermion mixtures in optical lattices is associated with the use of the hard-core boson approach,
where the occupation of on-site states conforms to the Pauli principle. For Bose atoms on the lattice, this model is a limiting case of Bose-Hubbard
model for $U\rightarrow\infty$ and is rather widely used \cite{mahan,micnas,seng2,ista2,ista3}. It is suitable for the region $0\leqslant \overline{n}_\textrm{B}\leqslant 1$,
but also can describe the MI-SF phase transitions in the vicinity of the points $\mu_\textrm{B}=nU$, $n=0,1,2,\ldots$ at finite values of $U$ (in the case of strong coupling, $t_\textrm{B}\ll U$) where $n\leqslant \overline{n}_\textrm{B} \leqslant n+1$ for $T=0$ within the SF phase region \cite{kona,ista1}. In these cases, the model is also applicable to these phase transitions at non-zero temperatures.

For boson-fermion mixtures, the BFH model in hard-core bosons limit remains less explored. In \cite{mys1},
the phase diagrams and phase separation or charge modulation conditions at the ion intercalation in semiconductor crystals were investigated
based on the BFH-type model; in \cite{mys2,mys3}, within pseudospin-fermion description (that corresponds to the above mentioned $U\rightarrow\infty$ limit)
the conditions of the appearance of SS and CDW phases under effective interparticle interactions were investigated.
For the four-state model, that arises in this case, the calculations for fermion band spectrum in Hubbard-I approximation were performed, and
its transformation during transition to the SF-phase and at the presence of a Bose-Einstein condensate \cite{kvo} was investigated.

The aim of this work is to more thoroughly study the thermodynamics of the above mentioned 4-state model. We confine ourselves to the case of ``heavy''
fermions (i.e., extremely low values of the fermion hopping parameter $t_\textrm{F}$). Such a situation was partially considered in \cite{mering,fehr2}.
It was argued, in particular, that frozen fermions, as fixed subsystem when $t_\textrm{F}=0$, are capable of preventing the occurrence of long-range correlations
of superfluid-type and the appearance of BE condensate. There exists, however, the critical fermion concentration below which this effect is absent
(for $d=2$, $\overline{n}^\textrm{crit}_{p}\sim 0.59$; for $d=3$, $\overline{n}^\textrm{crit}_{p}\sim 0.31$; see \cite{mering}).

We consider the equilibrium case, assuming that $t_\textrm{f}$ takes small values, but not so small that could lead to the state of a glass type \cite{bloch1,sanpera,shults}.
We shall use the mean-field approximation, basing, however, on an accurate allowance for interspecies interaction in the spirit of the strong coupling approach.
An analysis of the MI-SF phase transition, based on the conditions of thermodynamic equilibrium, will be performed
(we do not restrict ourselves to the criterion of the normal (MI) phase stability in order to determine the phase transition point). The study will be performed for the case of fixed chemical potentials of bosons $\mu_\textrm{B}$ and fermions $\mu_\textrm{F}$
in the zero temperature limit ($T=0$).
Corresponding phase diagrams will be built, taking into account the possibility of a change of the phase transition order from the second to the first order.
Our investigation will cover the case of a repulsive on-site interaction between hard-core bosons and fermions ($U_\textrm{BF} > 0$).

\section{The four-state model}
\label{sec:2}

The Hamiltonian of the Fermi-Bose-Hubbard model is usually written in the form:
\begin{flalign}
H=&\frac{U}{2}\sum\limits_{i}n_{i}^\textrm{b}(n_{i}^\textrm{b}-1)
+U'\sum\limits_{i}n_{i}^\textrm{b}n_{i}^\textrm{f}-\mu\sum\limits_{i}n_{i}^\textrm{b}-\mu'\sum\limits_{i}n_{i}^\textrm{f}
+\sum\limits_{\langle i,j \rangle} t_{ij}b^{+}_{i}b_{j}+\sum\limits_{\langle i,j \rangle} t_{ij}'a^{+}_{i}a_{j}.
\end{flalign}
Here, $U$ and $U'$ are constants of boson-boson and boson-fermion on-site interactions; $\mu$ and $\mu'$ are chemical potentials of bosons and fermions, respectively
(we consider here the case of repulsive interactions $U>0$, $U'>0$)
and $t$, $t'$ are tunneling amplitudes of bosons (fermions) describing the boson (fermion) hopping between the nearest lattice sites.

Let us use, as in \cite{kras}, the single-site basis of states
\begin{flalign}
&(n_{i}^\textrm{b}=n; \ n_{i}^\textrm{f}=0)\equiv|n,i\rangle, \qquad (n_{i}^\textrm{b}=n; \ n_{i}^\textrm{f}=1)\equiv|\widetilde{n},i\rangle,
\end{flalign}
where $n_{i}^\textrm{b}(n_{i}^\textrm{f})$ are occupation numbers of bosons (fermions) on the site $i$, and introduce the Hubbard operators
$X_{i}^{n,m}=|n,i\rangle\langle m,i|$, $X_{i}^{\widetilde{n},\widetilde{m}}=|\widetilde{n},i\rangle\langle \widetilde{m},i|$, etc.
Creation and destruction operators as well as operators of occupation numbers will be expressed in terms of $X$-operators in the following way \cite{ista1,kras}:
\begin{flalign}
b_i&=\sum\limits_n\sqrt{n+1}X_i^{n,n+1}
+\sum\limits_{\widetilde{n}}\sqrt{\widetilde{n}+1}X_i^{\widetilde{n},\widetilde{n}+1},\nonumber\\
b^+_i&=\sum\limits_n\sqrt{n+1}X_i^{n+1,n}
+\sum\limits_{\widetilde{n}}\sqrt{\widetilde{n}+1}X_i^{\widetilde{n}+1,\widetilde{n}},\nonumber\\
a_i&=\sum\limits_n X_i^{n,\widetilde{n}},\, \qquad a_i^+=\sum\limits_n X_i^{\widetilde{n},n},\nonumber\\
n_i^\textrm{b}&=\sum\limits_n nX^{n,n}+\sum\limits_{\widetilde{n}} \widetilde{n}X^{\widetilde{n},\widetilde{n}},\qquad n_i^\textrm{f}=\sum\limits_{\widetilde{n}} X^{\widetilde{n},\widetilde{n}},
\end{flalign}

Then, the Hamiltonian in this new representation takes the form:
\begin{flalign}\label{eq:ham24}
H&=H_0+H_1^\textrm{b}+H_1^\textrm{f},\\
H_0&=\sum\limits_{i,n}\lambda_{n}X_i^{nn}
+\sum\limits_{i,\widetilde{n}}\lambda_{\widetilde{n}}X_i^{\widetilde{n}\widetilde{n}},\nonumber\\
\lambda_{n}&=\frac{U}{2}n(n-1)-n\mu, \qquad  \lambda_{\widetilde{n}}=\frac{U}{2}\widetilde{n}(\widetilde{n}-1)-\mu\widetilde{n}-\mu'+U'\widetilde{n},\nonumber\\
H_1^\textrm{b}&=\sum\limits_{\langle i,j \rangle} t_{ij}b^{+}_{i}b_{j}, \qquad \qquad H_1^\textrm{f}=\sum\limits_{\langle i,j \rangle} t_{ij}'a^{+}_{i}a_{j}.\nonumber
\end{flalign}

In the case of a hard-core boson approximation $(U\rightarrow\infty)$, the single-site $|n_i^\textrm{b},n_i^\textrm{f}\rangle$ basis consists of four states:
\begin{flalign}
&|0\rangle=|0,0\rangle, \quad \quad |\widetilde{0}\rangle=|0,1\rangle, \nonumber\\
&|1\rangle=|1,0\rangle, \quad \quad |\widetilde{1}\rangle=|1,1\rangle.
\end{flalign}

In this limit,
\begin{align}
b_i&=X^{01}_i+X^{\widetilde{0}\widetilde{1}}, & b_i^+&=X^{10}_i+X^{\widetilde{1}\widetilde{0}}, \nonumber\\
a_i&=X^{0\widetilde{0}}_i+X^{1\widetilde{1}}, & a_i^+&=X^{\widetilde{0}0}_i+X^{\widetilde{1}1}, \\
n_i^\textrm{b}&=X_i^{11}+X_i^{\widetilde{1}\widetilde{1}}, & n_i^\textrm{f}&=X_i^{\widetilde{0}\widetilde{0}}+X_i^{\widetilde{1}\widetilde{1}},\nonumber
\end{align}
\begin{flalign}
&\lambda_0=0, \qquad \lambda_1=-\mu, \qquad \lambda_{\widetilde{0}}=-\mu', \qquad \lambda_{\widetilde{1}}=-\mu-\mu'+U'
\end{flalign}
and in the expression (\ref{eq:ham24}) for Hamiltonian of the system, a restriction $n=0,1$ and $\widetilde{n}=\widetilde{0},\widetilde{1}$ on occupation numbers is imposed.
As it was mentioned, we consider the case of the so-called heavy fermions, when the inequalities $t'\ll t$, $t'\ll U'$ are fulfilled. Our aim consists in the study of
conditions at which the MI-SF transition in such a model takes place, for the case when the fermion hopping between lattice sites can be neglected. On this
assumption, we shall start with Hamiltonian:
\begin{flalign}
&\hat{H}=\sum\limits_i \left(\lambda_{0}X_i^{00}+\lambda_{1}X_i^{11}+\lambda_{\widetilde{0}}X_i^{\widetilde{0}\widetilde{0}}+\lambda_{\widetilde{1}}X_i^{\widetilde{1}\widetilde{1}}\right)
+\sum\limits_{\langle i,j\rangle} t_{ij}b_i^+b_j .
\end{flalign}

\section{Mean field approximation}
\label{sec:3}

Let us introduce the order parameter of BE condensate $\varphi=\langle b_i\rangle = \langle b_i^+\rangle$. In the case of mean field approximation (MFA):
\begin{flalign}
&b_i^+b_j\rightarrow \varphi(b_i^++b_i)-\varphi^2, \nonumber\\
&\sum\limits_{ij}t_{ij}b_i^+b_j=\varphi t_0\sum\limits_i(b_i^++b_i)-Nt_0\varphi^2,
\end{flalign}
(here, $t_0=\sum t_{ij}=-|t_0|$, $t_0<0$).

Then, for initial Hamiltonian after separating the mean field part we shall have:
\begin{flalign}
&H=H_\textrm{MF}+\sum\limits_{i,j}t_{ij}(b_i^+-\varphi)(b_i-\varphi).
\end{flalign}

Here,
\begin{flalign}
&H_\textrm{MF}=\sum_i H_i-Nt_0\varphi^2, \qquad H_i=\sum\limits_{pr}H_{pr}X^{pr}_i,
\end{flalign}
and
\begin{flalign}
&||H_{pr}||=\left(
  \begin{array}{cccc|c}
    |0\rangle & |1\rangle & |\widetilde{0}\rangle & |\widetilde{1}\rangle &  \\
    \hline
    0 & t_0\varphi & 0 & 0 & |0\rangle \\
    t_0\varphi & -\mu & 0 & 0 & |1\rangle \\
    0 & 0 & -\mu' & t_0\varphi & |\widetilde{0}\rangle \\
    0 & 0 & t_0\varphi & -\mu-\mu'+U'& |\widetilde{1}\rangle \\
  \end{array}
\right).\end{flalign}

Our next step is diagonalization:
\begin{flalign}
&\hat{U}^T*\hat{H}*\hat{U}=\hat{\widetilde{H}},
\end{flalign}
where $\hat{U}=\left(
  \begin{array}{cc}
      \hat{U_1}& \hat{0} \\
        \hat{0} & \hat{U_2} \\
     \end{array}
\right)$
and
$\hat{U}_1=\left(
   \begin{array}{cc}
     \cos{\psi} & -\sin{\psi} \\
     \sin{\psi} & \cos{\psi} \\
   \end{array}
 \right)$,
$\hat{U}_2=\left(
   \begin{array}{cc}
     \cos{\widetilde{\psi}} & -\sin{\widetilde{\psi}} \\
     \sin{\widetilde{\psi}} & \cos{\widetilde{\psi}} \\
   \end{array}
 \right)
$.
Here,
\begin{equation}\label{eq:sin}
\sin{2\psi}=\frac{t_0\varphi}{\sqrt{\mu^2/4+t^2_0\varphi^2}}\,, \qquad \sin{2\widetilde{\psi}}=\frac{t_0\varphi}{\sqrt{(U'-\mu)^2/4+t^2_0\varphi^2}}\,.
\end{equation}

Then we will get a diagonal single-site part (which is also a mean-field part) of the Hamiltonian:
\begin{eqnarray}
\hat{H}_i=\sum\limits_{p'}\varepsilon_{p'} X_i^{p'p'}-Nt_0\varphi^2,
\end{eqnarray}
where $p'=0',1',\widetilde{0'},\widetilde{1'}$ are indices which denote the states of a new basis,
\begin{flalign}\label{eq:enlev1}
&\varepsilon_{0',1'}=-\frac{\mu}{2}\pm\sqrt{\frac{\mu^2}{4}+t_0^2\varphi^2}\,,\nonumber\\
&\varepsilon_{\widetilde{0'},\widetilde{1'}}=-\mu'-\frac{\mu}{2}
+\frac{U'}{2}\pm\sqrt{\frac{(U'-\mu)^2}{4}+t_0^2\varphi^2}\,.
\end{flalign}

For Bose-operators,  in a new basis we shall have:
\begin{flalign}
b_i=&\frac{1}{2}\sin(2\psi)(X_i^{0'0'}-X_i^{1'1'})
+\frac{1}{2}\sin(2\widetilde{\psi})(X_i^{\widetilde{0'}\widetilde{0'}}
-X_i^{\widetilde{1'}\widetilde{1'}})+\nonumber\\
&+\cos^2{\psi}X^{0'1'}_i-\sin^2{\psi}X^{1'0'}_i
+\cos^2{\widetilde{\psi}}X^{\widetilde{0'}\widetilde{1'}}_i
-\sin^2{\widetilde{\psi}}X^{\widetilde{1'}\widetilde{0'}}_i.\nonumber
\end{flalign}

\section{Grand canonical potential and order parameter}
\label{sec:4}

The partition function in MFA is equal to:
\begin{flalign}
Z_\textrm{MF}&=\textrm{Sp} \, \re^{-\beta H_\textrm{MF}}=\re^{\beta Nt_0\varphi^2}\prod\limits_i \textrm{Sp}\left\{ \exp\left(-\beta\sum\limits_{pr}H_{pr}X^{pr}_i\right)\right\}\nonumber\\
&=\re^{\beta Nt_0\varphi^2}\prod\limits_i \exp\left(-\beta\sum\limits_{p'}\varepsilon_{p'} X_i^{p'p'}\right)=\re^{\beta Nt_0\varphi^2}Z_0^N\,,
\end{flalign}
where
\begin{flalign}
&Z_0=\re^{-\beta\varepsilon_{0'}}+\re^{-\beta\varepsilon_{1'}}
+\re^{-\beta\varepsilon_{\widetilde{0'}}}+\re^{-\beta\varepsilon_{\widetilde{1'}}}.
\end{flalign}

The grand canonical potential is:
\begin{flalign}
&\Omega_\textrm{MF}=-\theta \ln{Z_\textrm{MF}}=N|t_0|\varphi^2-N\Theta \ln{Z_0}
\end{flalign}
or
\begin{flalign}
&\Omega_\textrm{MF}/N=|t_0|\varphi^2-\theta \ln{Z_0}
\end{flalign}
(here, we take into account that $t_0=-|t_0|$).
The equilibrium value of the order parameter $\varphi$ can be found from the global minimum condition of $\Omega$.

We have an equation
\begin{flalign}
&\frac{\partial(\Omega_\textrm{MF}/N)}{\partial \varphi}=2|t_0|\varphi-\frac{\theta}{Z_0}\frac{\partial Z_0}{\partial \varphi}=0
\end{flalign}
or
\begin{flalign}\label{eq:4.6}
&2|t_0|\varphi+\sum\limits_{p'}\langle X^{p'p'}\rangle\frac{\partial \varepsilon_{p'}}{\partial \varphi}=0.
\end{flalign}
Here:
\begin{flalign}
&\langle X^{p'p'}\rangle=\frac{1}{Z_0}\re^{-\beta\varepsilon_{p'}}
\end{flalign}
Using that:
\begin{flalign}
&\frac{\partial \varepsilon_{0',1'}}{\partial \varphi}=\pm t_0\sin{2\psi}=\mp|t_0|\sin{2\psi},\nonumber\\
&\frac{\partial \varepsilon_{\widetilde{0'},\widetilde{1'}}}{\partial \varphi}=\pm t_0\sin{2\widetilde{\psi}}=\mp|t_0|\sin{2\widetilde{\psi}},
\end{flalign}
from (\ref{eq:4.6})  we shall get:
\begin{flalign}
&\varphi=\frac{1}{2}\sin{2\psi}\left(\langle X^{0'0'} \rangle-\langle X^{1'1'} \rangle\right)+
\frac{1}{2}\sin{2\widetilde{\psi}}\left(\langle X^{\widetilde{0'}\widetilde{0'}} \rangle-\langle X^{\widetilde{1'}\widetilde{1'}} \rangle\right)
\end{flalign}
or in the explicit form,
\begin{flalign}
&\varphi=\frac{|t_0|\varphi}{2} \left(\frac{\langle X^{1'1'} \rangle-\langle X^{0'0'} \rangle}{\sqrt{\frac{\mu^2}{4}+t_0^2\varphi^2}}+
\frac{\langle X^{\widetilde{1'}\widetilde{1'}} \rangle-\langle X^{\widetilde{0'}\widetilde{0'}} \rangle}{\sqrt{\frac{(U'-\mu)^2}{4}+t_0^2\varphi^2}}\right).
\end{flalign}

This equation has trivial $\varphi=0$ and non-trivial $\varphi\neq 0$ solutions, the second one can be obtained from the equation:
\begin{flalign} \label{eq:main}
&\frac{1}{|t_0|}=\frac{\langle X^{1'1'} \rangle-\langle X^{0'0'} \rangle}{2\sqrt{\frac{\mu^2}{4}+t_0^2\varphi^2}}+
\frac{\langle X^{\widetilde{1'}\widetilde{1'}} \rangle-\langle X^{\widetilde{0'}\widetilde{0'}} \rangle}{2\sqrt{\frac{(U'-\mu)^2}{4}+t_0^2\varphi^2}}\,.
\end{flalign}

When we have several solutions in this equation, we shall consider only those related to the minimum of $\Omega_\textrm{MF}$.

Let us apply the unitary transformation $\hat{U}^{T}(\ldots)\hat{U}$ to operators $X_i^{01}$ and $X_i^{\widetilde{0}\widetilde{1}}$, which, in the matrix form, are:
\begin{flalign}
&||X_i^{01}||=\left(
  \begin{array}{cccc}
    0 & 1 & 0 & 0\\
     0 & 0 & 0 & 0\\
      0 & 0 & 0 & 0\\
       0 & 0 & 0 & 0
  \end{array}
\right), \qquad
||X_i^{\widetilde{0}\widetilde{1}}||=\left(
  \begin{array}{cccc}
    0 & 0 & 0 & 0\\
     0 & 0 & 0 & 0\\
      0 & 0 & 0 & 1\\
       0 & 0 & 0 & 0
  \end{array}
\right).
\end{flalign}
We shall get:
\begin{flalign}
&||\hat{U}^{T}X_i^{01}\hat{U}||=\left(
  \begin{array}{cccc}
    \sin{\psi}\cos{\psi} & \cos^2{\psi} & 0 & 0\\
     -\sin^2{\psi} & -\sin{\psi}\cos{\psi} & 0 & 0\\
      0 & 0 & 0 & 0\\
       0 & 0 & 0 & 0
  \end{array}
\right)\end{flalign}
 and
 \begin{flalign}
&||\hat{U}^{T}X_i^{\widetilde{0}\widetilde{1}}\hat{U}||=\left(
  \begin{array}{cccc}
       0 & 0 & 0 & 0\\
       0 & 0 & 0 & 0\\
       0 & 0 & \sin{\widetilde{\psi}}\cos{\widetilde{\psi}}&  \cos^2{\widetilde{\psi}}\\
       0 & 0 & -\sin^2{\widetilde{\psi}} & -\sin{\widetilde{\psi}}\cos{\widetilde{\psi}}
  \end{array}
\right).\end{flalign}

In the transformed basis representation:
\begin{flalign}
&\hat{U}^{T}X_i^{01}\hat{U}=\cos^2{\psi}X_i^{0'1'}+\sin{\psi}\cos{\psi}(X_i^{0'0'}-X_i^{1'1'})
-\sin^2{\psi}X_i^{1'0'},\nonumber\\
&\hat{U}^{T}X_i^{\widetilde{0}\widetilde{1}}\hat{U}=\cos^2{\widetilde{\psi}}X^{\widetilde{0'}\widetilde{1'}}+\sin{\widetilde{\psi}}\cos{\widetilde{\psi}}( X^{\widetilde{0'}\widetilde{0'}}-X^{\widetilde{1'}\widetilde{1'}})
-\sin^2{\widetilde{\psi}}X^{\widetilde{1'}\widetilde{0'}}.
\end{flalign}

After averaging with the aid of $H_\textrm{MF}$ Hamiltonian:
\begin{flalign}
&\langle X_i^{01}\rangle=\frac{1}{Z_0}\textrm{Sp} \left(X_i^{01}\re^{-\beta H_\textrm{MF}}\right)=\frac{1}{Z_0} \textrm{Sp}\left(\hat{U}^{T}X_i^{01}\hat{U}e^{-\beta H_\textrm{MF}}\right).
\end{flalign}

On the new basis, the Hamiltonian $H_\textrm{MF}$ is diagonal; that is why the averages of diagonal operators $X_i^{p'p'}$ ($p'=0',1',\widetilde{0'},\widetilde{1'}$) will only be non-zero. Thus,
\begin{flalign}
&\langle X_i^{01}\rangle=\frac{1}{2}\sin{2\psi}\left(\langle X^{0'0'}\rangle-\langle X^{1'1'}\rangle\right),\nonumber\\
&\langle X_i^{\widetilde{0}\widetilde{1}}\rangle=\frac{1}{2}\sin{2\widetilde{\psi}}\left(\langle X^{\widetilde{0'}\widetilde{0'}}\rangle-\langle X^{\widetilde{1'}\widetilde{1'}}\rangle\right).
\end{flalign}
As a result, we have:
\begin{flalign}
\varphi=\langle b_i\rangle=\langle  X_i^{01}\rangle+\langle X_i^{\widetilde{0}\widetilde{1}}\rangle=\frac{1}{2}\sin{2\psi}\left(\langle X^{0'0'}\rangle-\langle X^{1'1'}\rangle\right)+
\frac{1}{2}\sin{2\widetilde{\psi}}\left(\langle X^{\widetilde{0'}\widetilde{0'}}\rangle-\langle X^{\widetilde{1'}\widetilde{1'}}\rangle\right).
\end{flalign}

Thus, we arrived at the same equation for $\varphi$ as we have got from the extremum condition for grand canonical potential.

\section{Spinodals at $T=0$}
\label{sec:5}

If in the equation (\ref{eq:main}) we substitute $\varphi=0$, we shall have a condition for the second order phase transition to SF phase
(if this transition is possible). In general, it is the condition of instability of normal (MI) phase with respect to the Bose-Einstein condensate appearance
(in figures it corresponds to spinodal lines).

The equation (\ref{eq:main}) can be rewritten as:
\begin{flalign}\label{eq:small}
&\frac{1}{|t_0|}=\frac{\langle X^{1'1'}\rangle-\langle X^{0'0'}\rangle}{\varepsilon_{0'}-\varepsilon_{1'}}+
\frac{\langle X^{\widetilde{1'}\widetilde{1'}} \rangle-\langle X^{\widetilde{0'}\widetilde{0'}} \rangle}{\varepsilon_{\widetilde{0'}}-\varepsilon_{\widetilde{1'}}}.
\end{flalign}

If $\varphi\rightarrow 0$,
\begin{flalign}
&\varepsilon_{0'}=\begin{cases}
\lambda_0\,, \ \mu > 0,\\
\lambda_1\,, \ \mu < 0; \end{cases}\qquad
\varepsilon_{\widetilde{0}}=\begin{cases}
\lambda_{\widetilde{1}}\,, \ \mu <U',\\
\lambda_{\widetilde{0}}\,, \ \mu > U'; \end{cases}\\
&\varepsilon_{1'}=\begin{cases}
\lambda_1\,, \ \mu > 0,\\
\lambda_0\,, \ \mu < 0; \end{cases} \qquad
\varepsilon_{\widetilde{1}}=\begin{cases}
\lambda_{\widetilde{0}}\,, \ \mu <U',\\
\lambda_{\widetilde{1}}\,, \ \mu > U'. \end{cases}\nonumber
\end{flalign}

 It is seen that we can divide the $\mu$ axis into three regions (1) $\mu <0$; (2) $0<\mu <U'$; (3) $\mu >U'$ (when $U'>0$).
 For all these regions, the equation (\ref{eq:small}) takes the form:
\begin{flalign}\label{eq:nextsm}
&\frac{1}{|t_0|}=\frac{\langle X^{00}\rangle-\langle X^{11}\rangle}{\lambda_1-\lambda_0}+
\frac{\langle X^{\widetilde{0}\widetilde{0}} \rangle-\langle X^{\widetilde{1}\widetilde{1}} \rangle}{\lambda_{\widetilde{1}}-\lambda_{\widetilde{0}}}\,.
\end{flalign}

It can be rewritten as:
\begin{flalign}\label{eq:next}
&\frac{1}{|t_0|}=\frac{\langle X^{11}\rangle-\langle X^{00}\rangle}{\mu}+
\frac{\langle X^{\widetilde{0}\widetilde{0}} \rangle-\langle X^{\widetilde{1}\widetilde{1}} \rangle}{U'-\mu}\,.
\end{flalign}

The equation (\ref{eq:nextsm}) is the same as the one obtained in \cite{mys3} from the condition of divergence of the bosonic Green's function
(calculated in the random phase approximation) at $\omega=0, {\bf q}=0$. In this way, it is the condition of instability of the phase with
 $\varphi=0$. Therefore, the equation (\ref{eq:nextsm}) is an equation for spinodal line.

When $T=0$, averages $\langle X^{p'p'}\rangle$, $\langle X^{\widetilde{p'}\widetilde{p'}}\rangle $ are different from zero only for the lowest energy level. Here, the three cases can be separated out.

\begin{enumerate}
\item $\mu'<0$.

 Here, at $\mu<0$, the ground state is $|0\rangle$, and at $\mu>0$ the ground state is $|1\rangle$. Respectively,
 in the first one of these cases $\langle X^{00}\rangle=1$, while in the second one $\langle X^{11}\rangle=1$ (other averages are equal to zero).
 The equation (\ref{eq:next}) can be written now as:
\begin{flalign}
&\frac{1}{|t_0|}=
\left\{
  \begin{array}{ll}
     -\frac{1}{\mu}, & \hbox{$\mu < 0$}, \\[1ex]
    \phantom{-}\frac{1}{\mu}, & \hbox{$\mu > 0$}.
  \end{array}
\right.
\end{flalign}

It follows from here that:
\begin{flalign}
&\mu=
\left\{
  \begin{array}{ll}
    \phantom{-}t_0, & \hbox{$\mu > 0$}, \\[1ex]
    -t_0, & \hbox{$\mu < 0$}.
  \end{array}
\right.
\end{flalign}
This is the spinodal equation for $\mu'<0$

\item $\mu'>U'$.

The change of ground state takes place when $\mu=U'$. For $\mu<U'$, the state $|\widetilde{0}\rangle$ is the ground one, and for $\mu>U'$ it is the state $|\widetilde{1}\rangle$. Respectively, at $\mu<U'$, $\langle X^{\widetilde{0}\widetilde{0}}\rangle=1$ and at $\mu>U'$, $\langle X^{\widetilde{1}\widetilde{1}}\rangle=1$. The equation (\ref{eq:next}) takes now the form:
\begin{flalign}
&\frac{1}{|t_0|}=
\left\{
  \begin{array}{ll}
    \frac{1}{U'-\mu}, & \hbox{$\mu < U'$}, \\[1ex]
    \frac{1}{\mu-U'}, & \hbox{$\mu > U'$}.
  \end{array}
\right.
\end{flalign}
In this case, the following lines will be the lines of spinodal:
\begin{flalign}
&\begin{cases}\mu=U'+t_0, \quad \mu > U',\\
\mu=U'-t_0, \quad \mu < U'.\end{cases}
\end{flalign}

\item $0<\mu'<U'$.

The change of ground state takes place now when $\mu=\mu'$. For $\mu<\mu'$, such a state is $|\widetilde{0}\rangle$, and for $\mu>\mu'$ it is
the state $|1\rangle$. Respectively, at $\mu<\mu'$, $\langle X^{\widetilde{0}\widetilde{0}}\rangle=1$ and for $\mu>\mu'$, $\langle X^{11}\rangle=1$.
For spinodal we shall now have the equation:
\begin{flalign}
&\frac{1}{|t_0|}=\frac{\langle X^{11}\rangle}{\mu}+\frac{\langle X^{\widetilde{0}\widetilde{0}}\rangle}{U'-\mu}\,.
\end{flalign}

It follows from here that, $\mu=U'-|t_0|$ when $\mu<\mu'$, and $\mu=|t_0|$ when $\mu>\mu'$. In the case $|t_0|<U'$, the solution $\mu=\mu'$ also appears.
When $\mu'>U'/2$, it exists for $U'-\mu'<|t_0|<\mu'$, and at $\mu'<U'/2$ it exists for $\mu'<|t_0|<U'-\mu'$. If $|t_0|\geqslant U'$,  the solution $\mu=\mu'$  disappears.

\end{enumerate}

The lines of spinodales are broken and much differ for the cases $|t_0|\geqslant U'$, $U'>|t_0|>U'/2$, $|t_0|<U'/2$. As a result, the areas of absolute
 instability of the normal phase (calculated at $T=0$) possess different shapes. These areas are shown below [see $(\mu, \mu')$ diagrams in section~\ref{sec:8}].

\section{Phase transition of the first order to SF phase}
\label{sec:6}

Let us analyze now the dependences of the order parameter $\varphi$ and grand canonical potential $\Omega$ upon the chemical potential of bosons
at different values of temperature and $\mu'$ using the equation (\ref{eq:main}). In the limit $T\rightarrow 0$, only the averages related to the ground state contribute to the right-hand side of the equation.
With the change of $\varphi$, the ground state can be reconstructed and this makes the problem of determining the solutions for order parameter self-consistent.
This problem can have a simple analytical solution, when the states with fermion and without fermion are not mutually competitive. This is achieved when $\mu'$ takes the values which are off the $[0, U']$ interval.

When we have the negative $\mu'$ values, the state $|1'\rangle$ is the ground one and the equation (\ref{eq:main}) reduces in this case:
\begin{flalign}
&\frac{1}{|t_0|}=\frac{1}{2\sqrt{\frac{\mu^2}{4}+t_0^2\varphi^2}}\,.
\end{flalign}
Then:
\begin{flalign}\label{eq:simpl1}
&\varphi=\frac{1}{2}\sqrt{1-\mu^2/t_0^2}\,.
\end{flalign}

In a positive region, when $\mu'>U'$, the state $|\widetilde{1'}\rangle$ is the ground one; respectively, we have an equation:
\begin{flalign}
&\frac{1}{|t_0|}=\frac{1}{2\sqrt{\frac{(U'-\mu)^2}{4}+t_0^2\varphi^2}}\,.
\end{flalign}
In this case:
\begin{flalign}\label{eq:simpl2}
&\varphi=\frac{1}{2}\sqrt{1-(\mu-U')^2/t_0^2}\,.
\end{flalign}

\begin{figure}[!b]
    \centerline{
    {\includegraphics[width=0.43\textwidth]{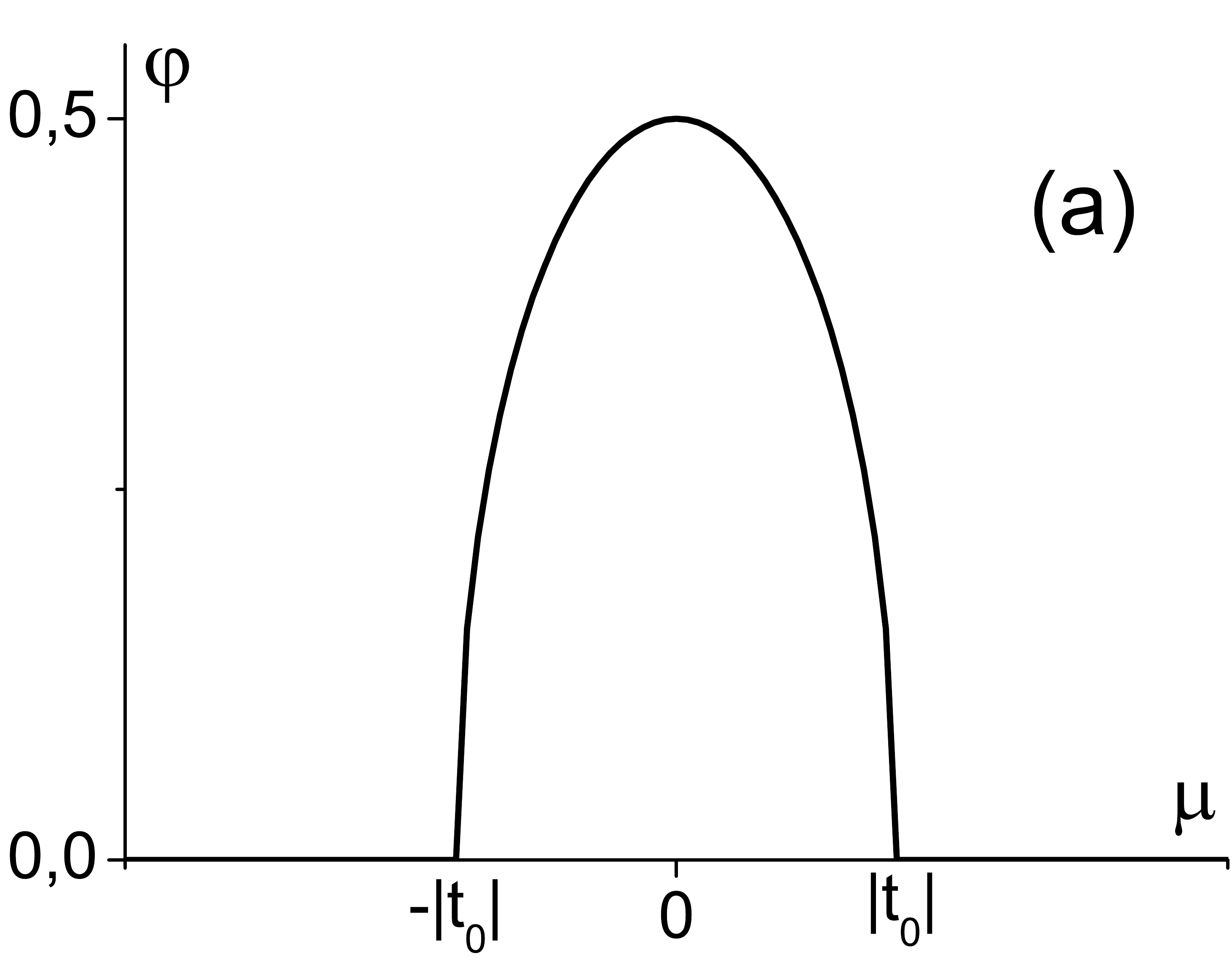}
        \label{fig:f1}}
        \hspace{5mm}
    {\includegraphics[width=0.43\textwidth]{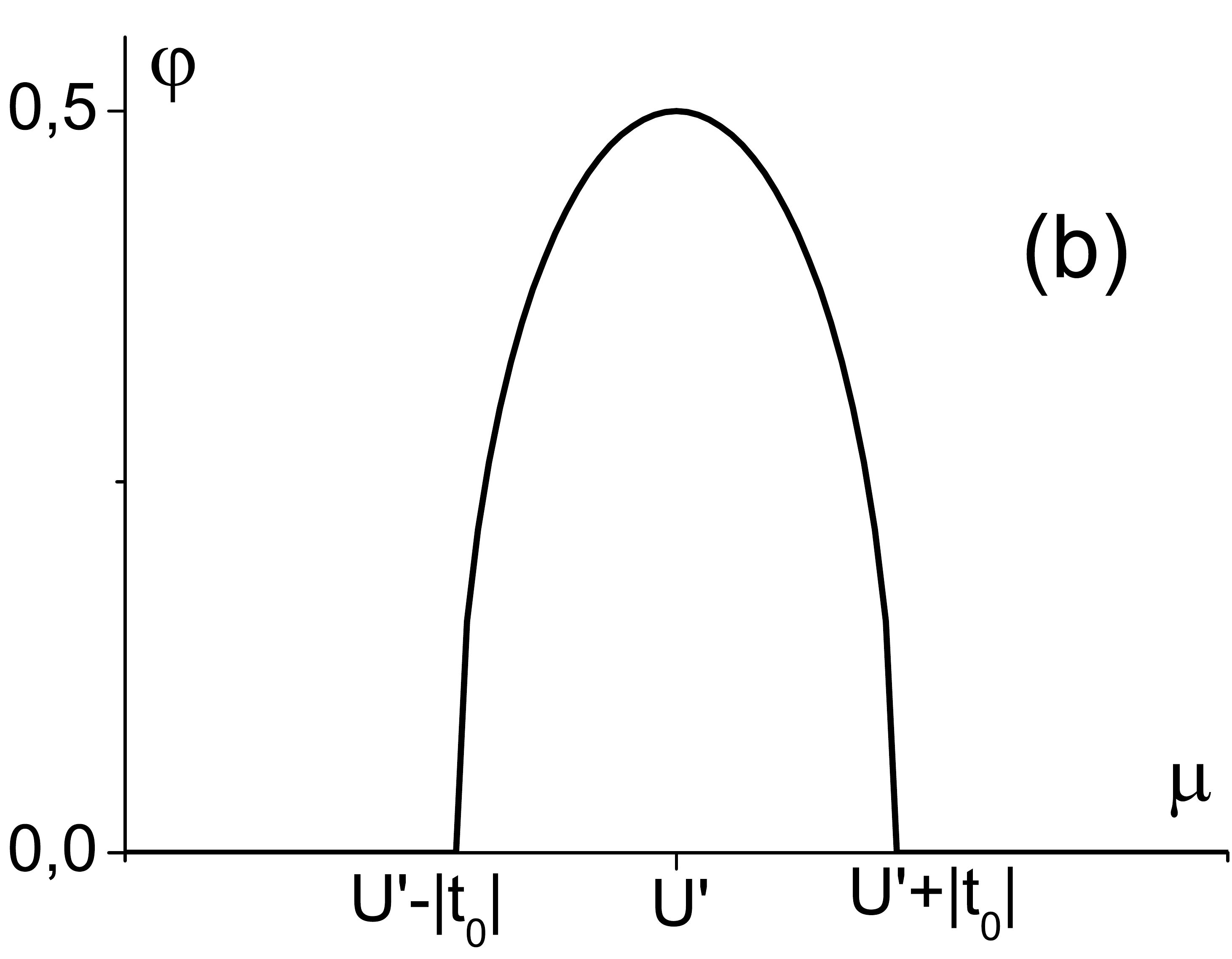}
        \label{fig:f2}}
        }
    \caption{Order parameter $\varphi$ as function of $\mu$ for $\mu'<0$ (a) and for $\mu'>U'$ (b).}
    \label{fig:fig1}
\end{figure}

The dependences of functions (\ref{eq:simpl1}) and (\ref{eq:simpl2}) on $\mu$ are presented in figure~\ref{fig:fig1}.

\begin{figure}[!t]
    \centering
    {\includegraphics[width=0.41\textwidth]{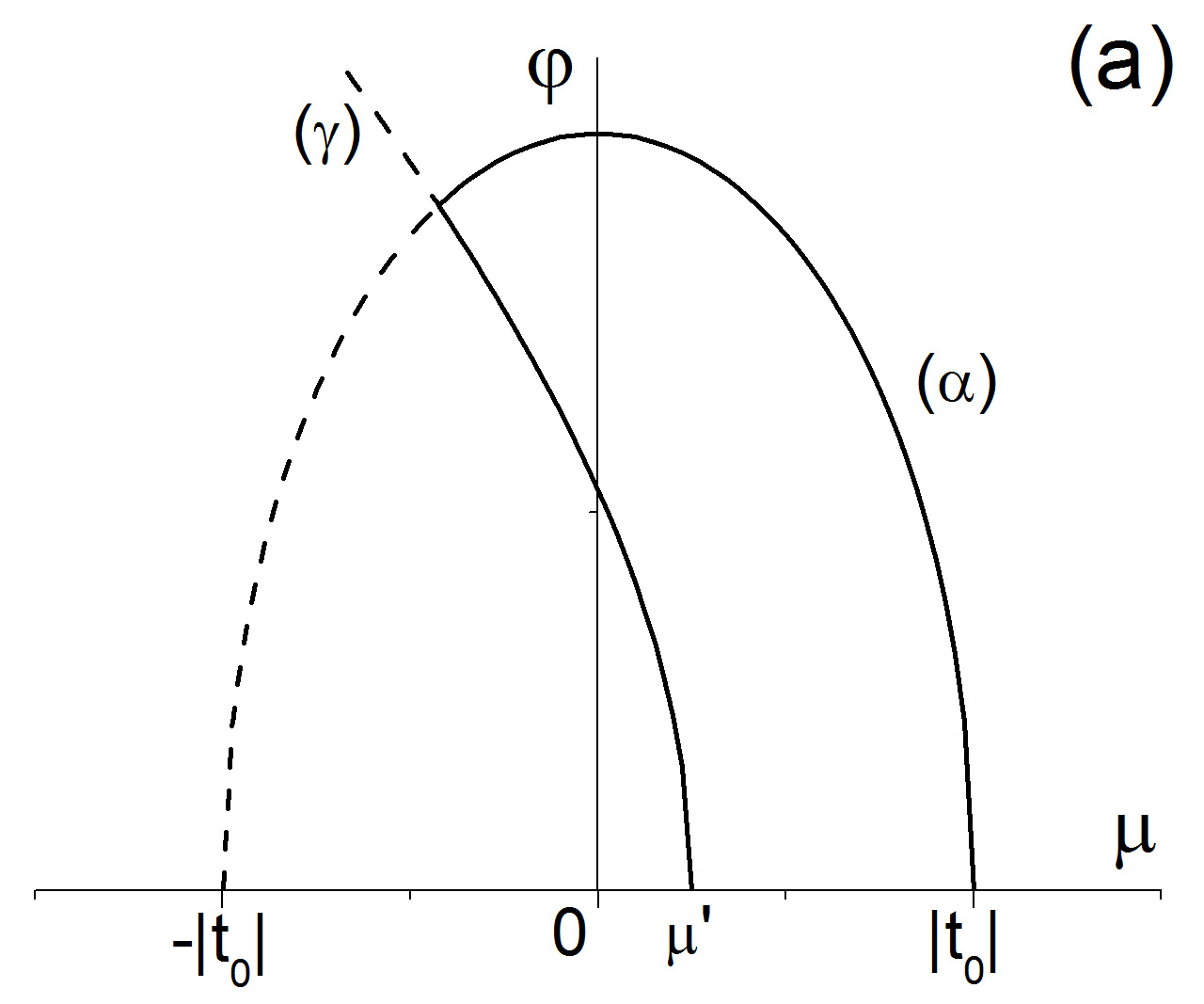}
        \label{fig:newfi1}}
\hspace{5mm}
    {\includegraphics[width=0.41\textwidth]{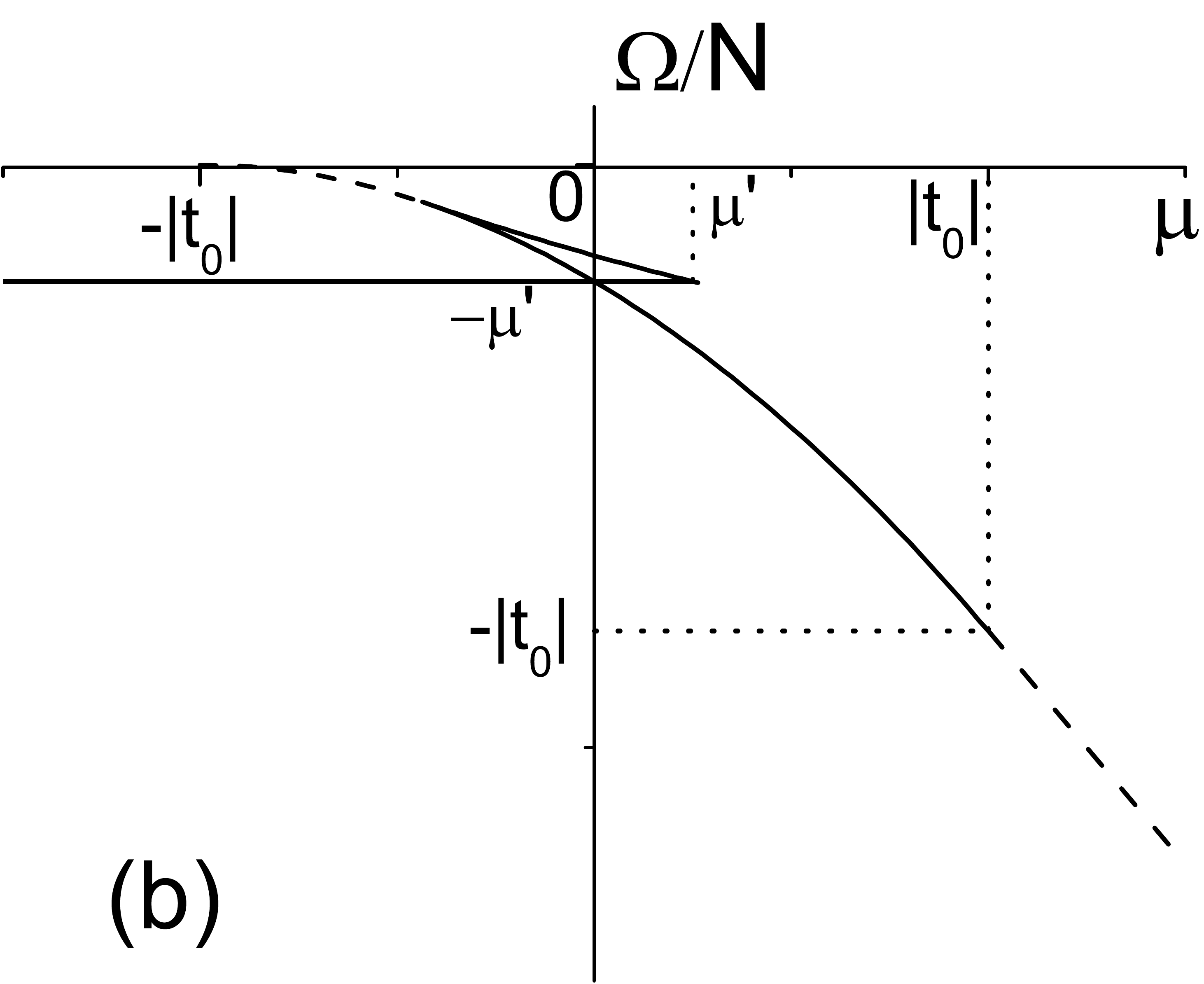}
        \label{fig:newom1}}
    \caption{Order parameter (a) and grand canonical potential (b) as functions of $\mu$ in the case $|t_0| <U'/2; \quad 0<\mu'<|t_0|$.
    Here, and in figures \ref{fig:new2}--\ref{fig:new5}, the lines $(\alpha)$, $(\beta)$ and $(\gamma)$ are described by formulae (\ref{eq:simpl1}), (\ref{eq:simpl2}) and
    (\ref{eq:simpl3}), respectively.}
    \label{fig:new1}
\end{figure}

\begin{figure}[!t]
    \centering
    {\includegraphics[width=0.41\textwidth]{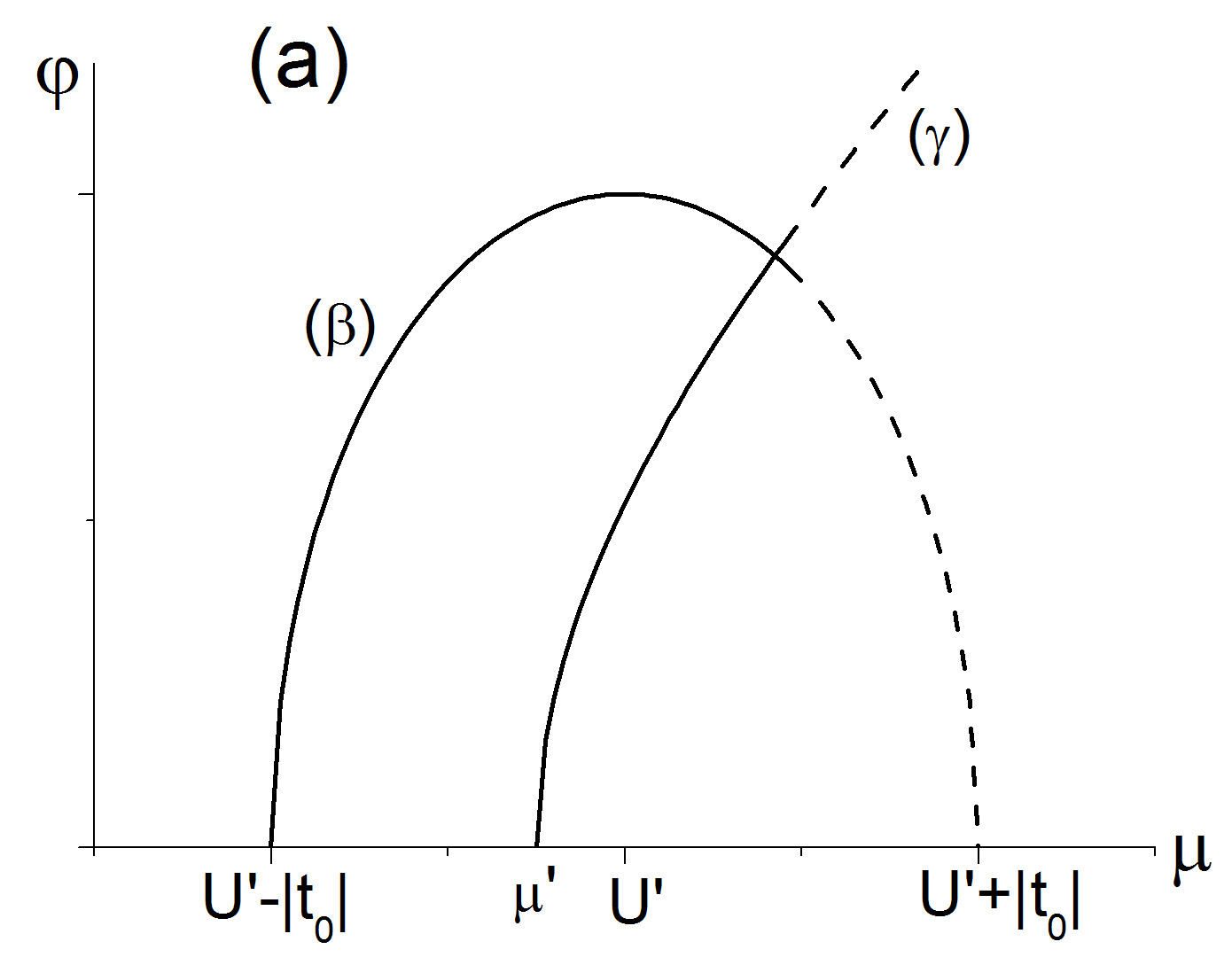}
        \label{fig:newfi2}}
\hspace{5mm}
    {\includegraphics[width=0.41\textwidth]{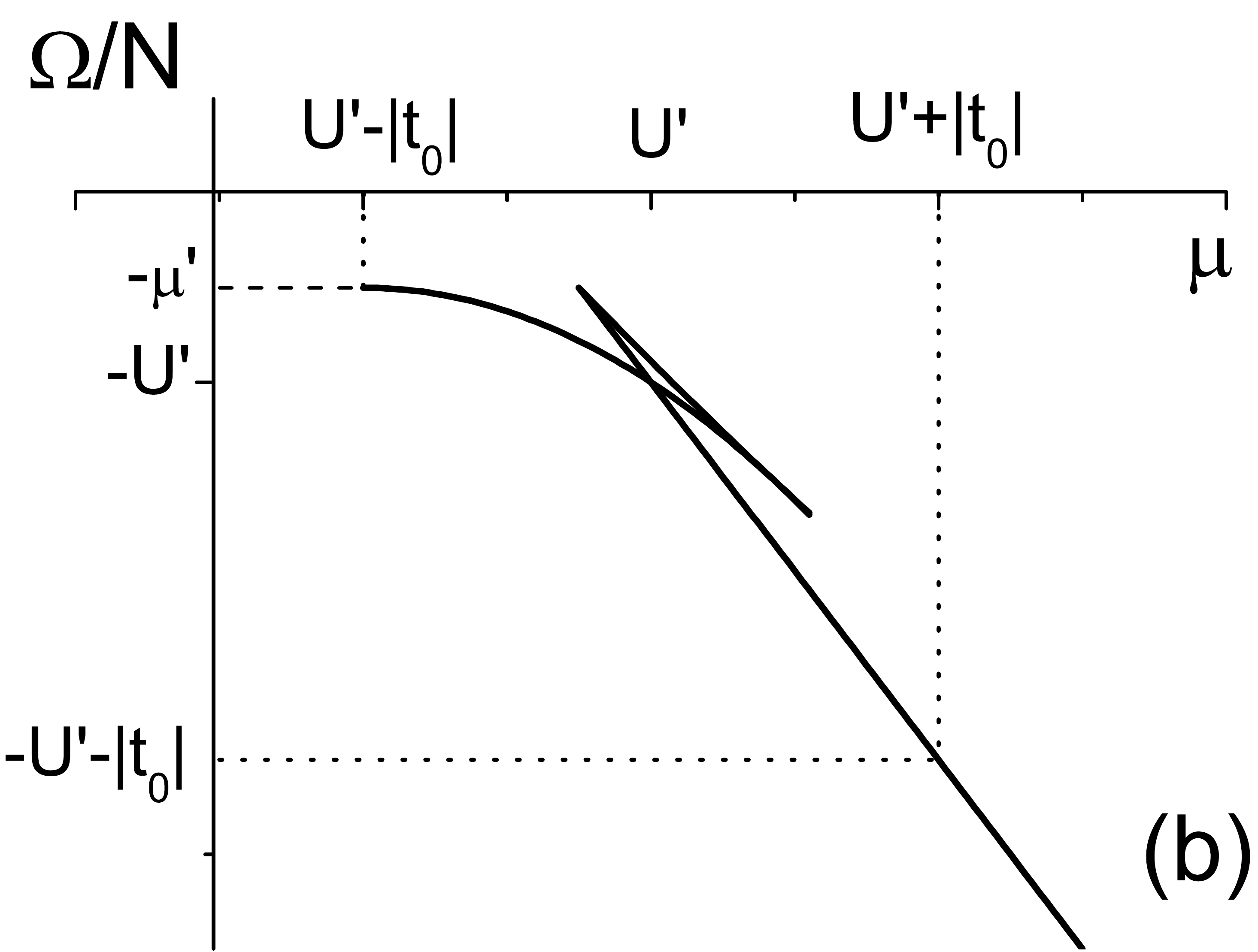}
        \label{fig:newom2}}
    \caption{Order parameter (a) and grand canonical potential (b) as functions of $\mu$ in the case $|t_0| <U'/2; \quad U'-|t_0|<\mu'<U'$.}
    \label{fig:new2}
\end{figure}

\begin{figure}[!t]
    \centering
    {\includegraphics[width=0.41\textwidth]{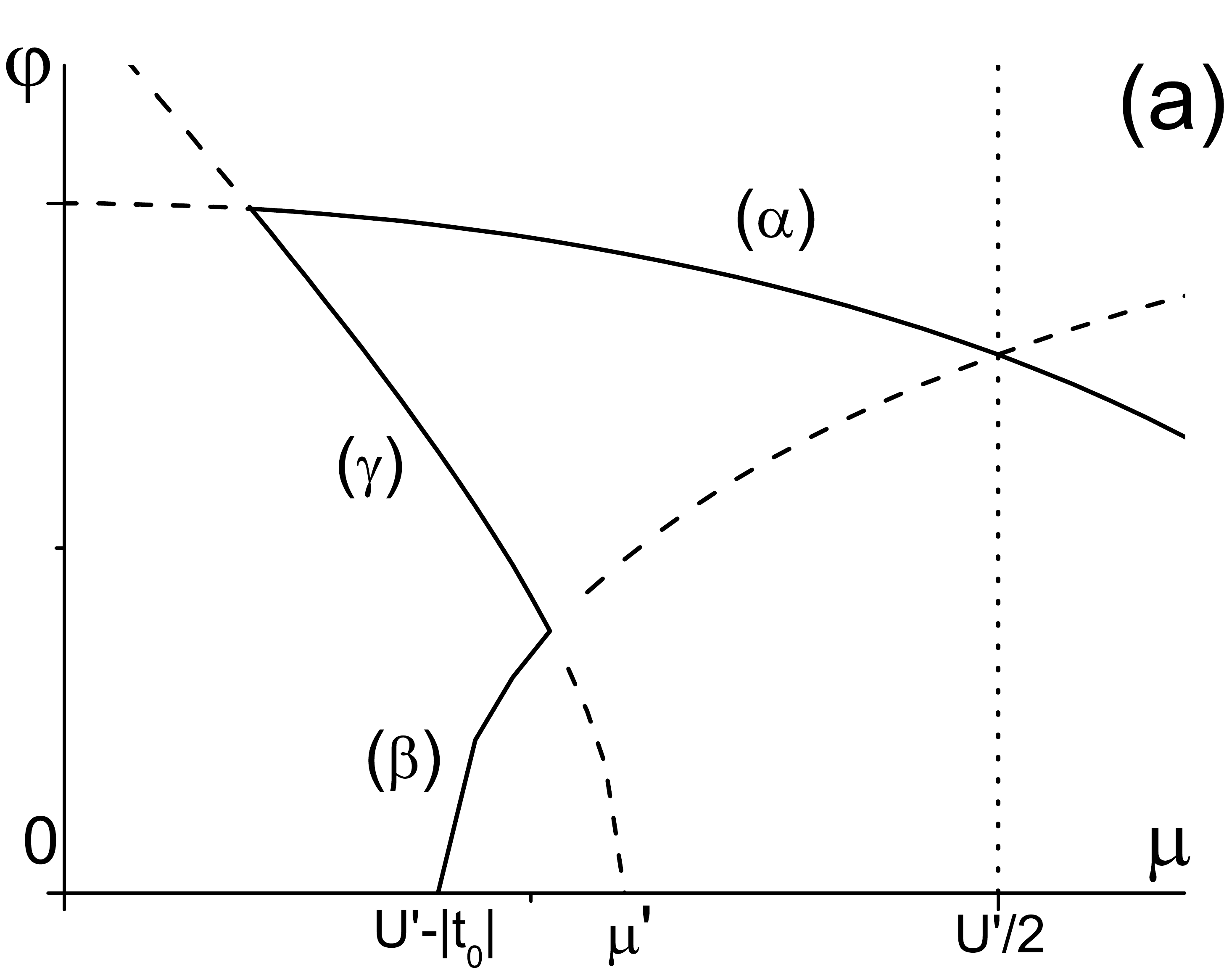}
        \label{fig:newfi3}}
\hspace{5mm}
    {\includegraphics[width=0.41\textwidth]{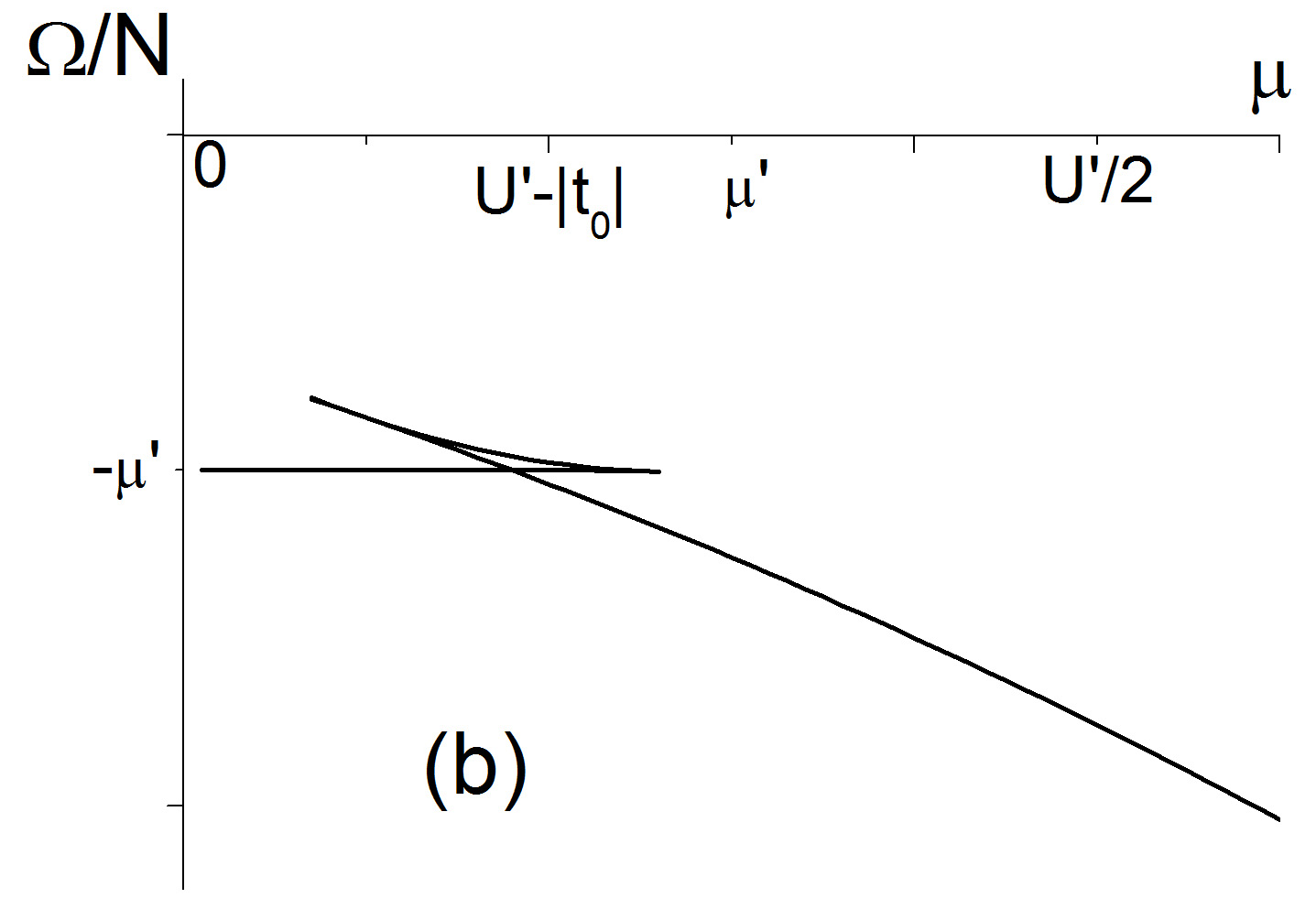}
        \label{fig:newom3}}
    \caption{Order parameter (a) and grand canonical potential (b) as functions of $\mu$ in the case $|t_0| > U'/2; \quad U'-|t_0|< \mu' < U'/2$.}
    \label{fig:new3}
\end{figure}

\begin{figure}[!t]
    \centering
    {\includegraphics[width=0.41\textwidth]{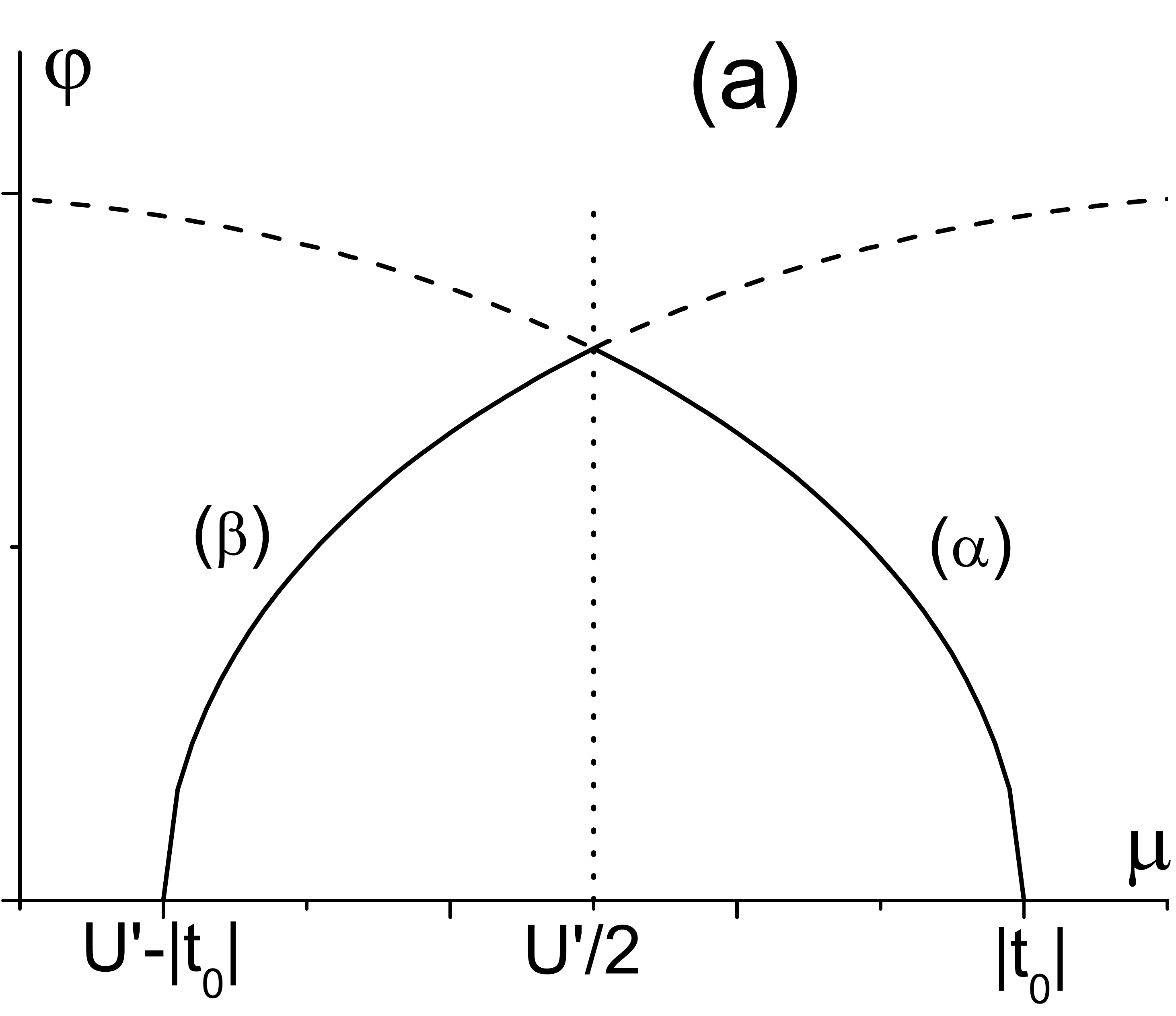}
        \label{fig:newfi4}}
\hspace{5mm}
    {\includegraphics[width=0.41\textwidth]{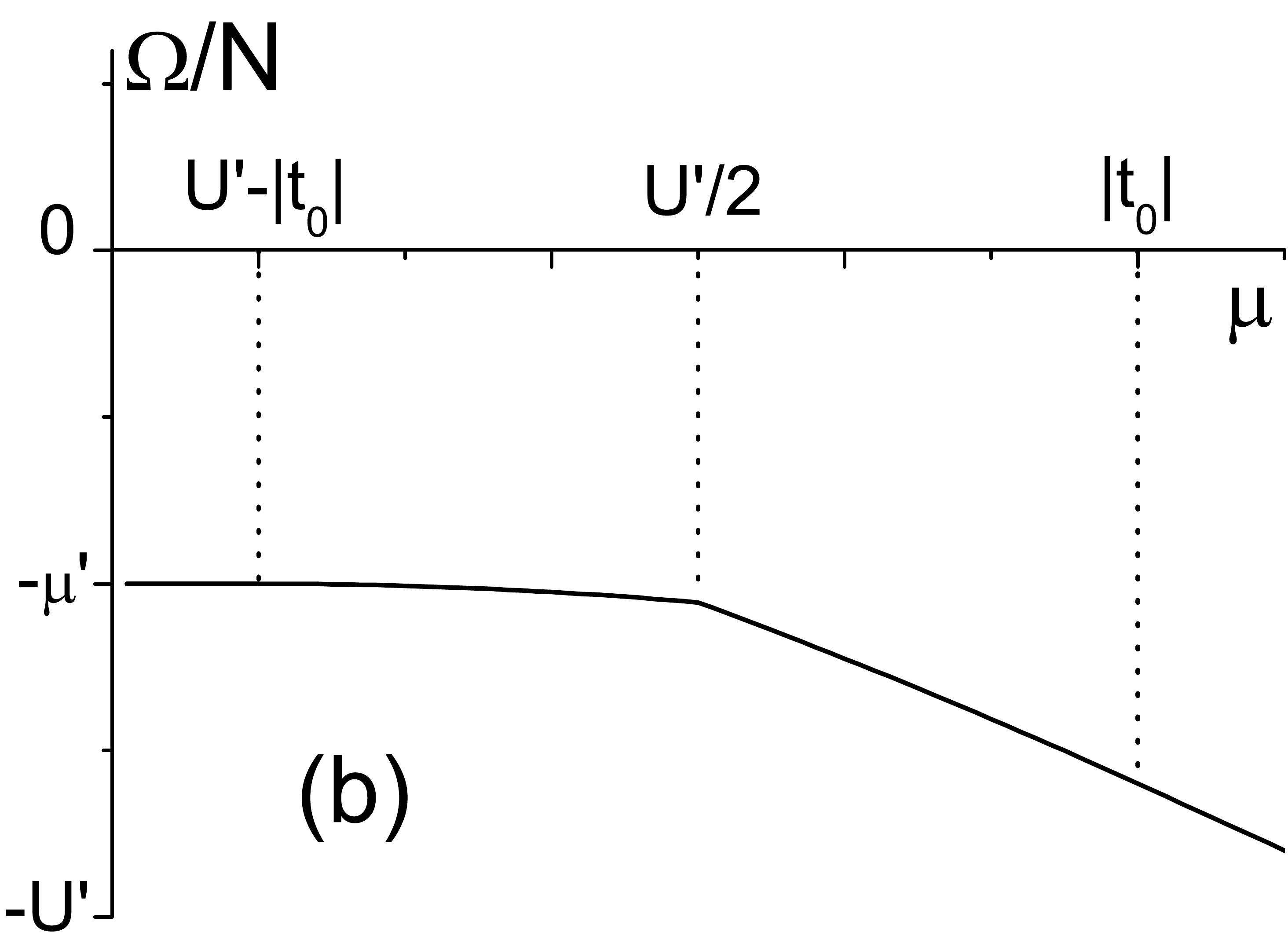}
        \label{fig:newom4}}
    \caption{Order parameter (a) and grand canonical potential (b) as functions of $\mu$ in the case when $|t_0| > U'/2; \quad \mu'=U'/2$.}
    \label{fig:new4}
\end{figure}

\begin{figure}[!t]
    \centering
    {\includegraphics[width=0.41\textwidth]{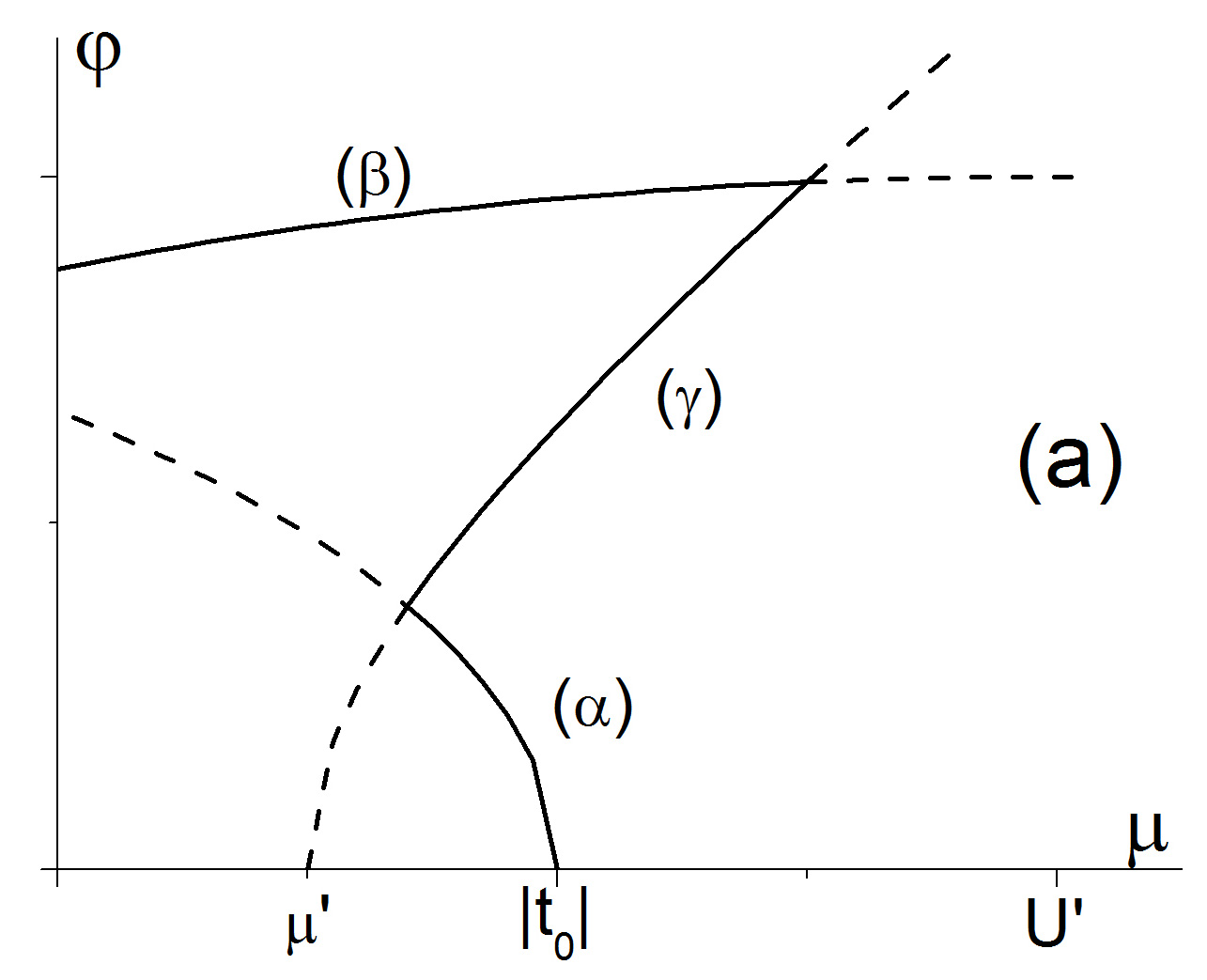}
        \label{fig:newfi5}}
\hspace{5mm}
    {\includegraphics[width=0.41\textwidth]{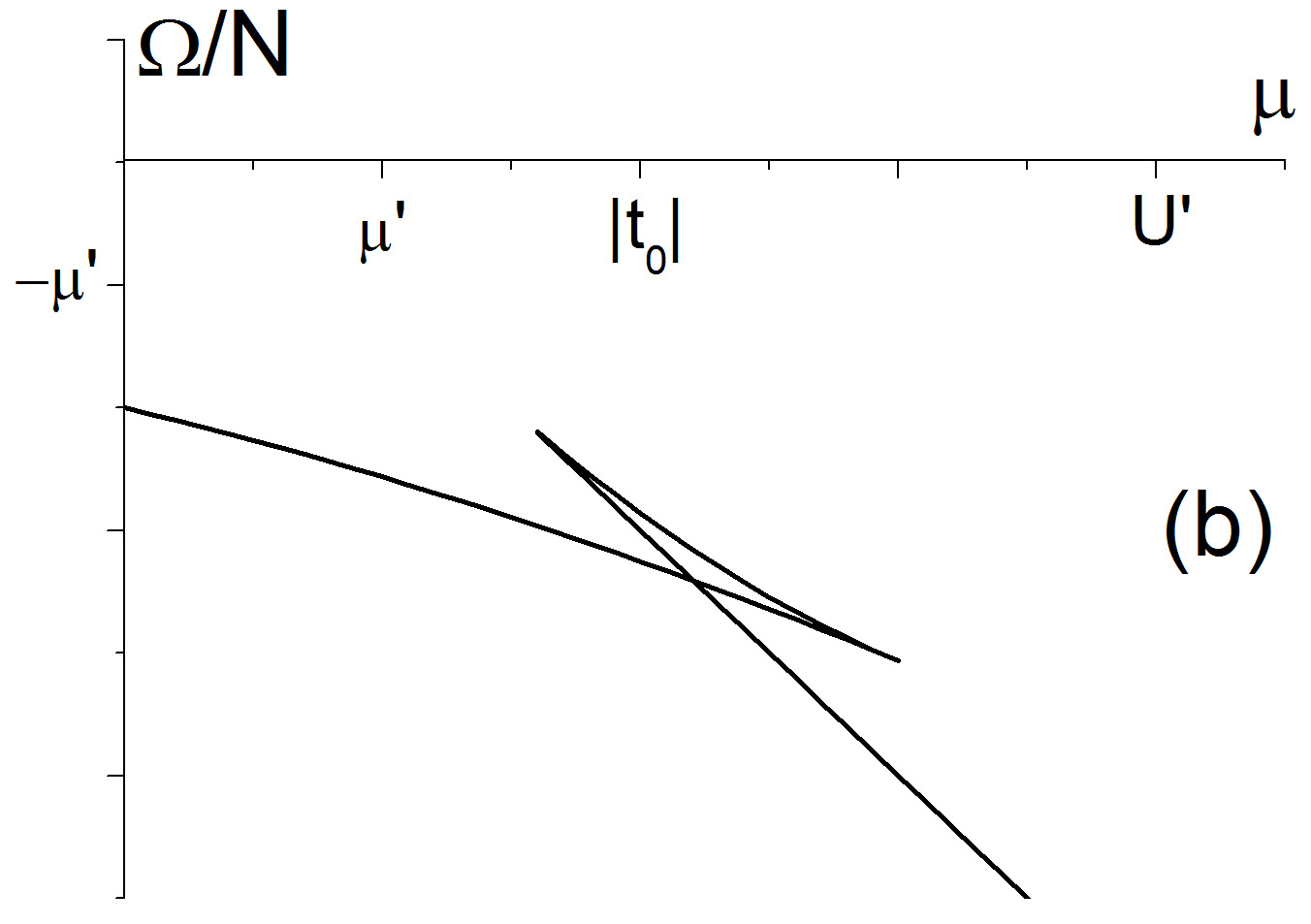}
        \label{fig:newom5}}
    \caption{Order parameter $\varphi$ and grand canonical potential $\Omega/N$ as function of $\mu$ in the case $|t_0| > U'/2; \quad U'/2<\mu'<|t_0|$.}
    \label{fig:new5}
\end{figure}

In the region of intermediate values of $\mu'$ (at $0\leqslant \mu'\leqslant U'$), the competition of ``tilded'' and ``untilded'' states leads to
deformation of the curve $\varphi(\mu)$. In figures \ref{fig:new1}~(a)--\ref{fig:new5}~(a) one can see the plots of the order parameter $\varphi$ as function of chemical potential of bosons $\mu$ for different values of chemical potential of fermions $\mu'$.  These curves are obtained numerically from the
equation (\ref{eq:main}) in the case $T=0$.

One can see that in intervals $0<\mu'<|t_0|$ and $U'-|t_0|<\mu'<U'$ (at $|t_0|<U'/2$) as well as in almost the whole interval $0<\mu'<U'$ (at $|t_0|>U'/2$)
of $\mu'$ values, the $\varphi(\mu)$ dependence has a reverse course and S-like behaviour. This is an evidence of the possibility of
the first order phase transition (rather than the second order transition).

This conclusion  can be confirmed by calculation of grand canonical potential $\Omega_\textrm{MF}(\mu)$ as function of~$\mu$.

In the case of negative values of chemical potential $\mu'$, when at $T=0$, only the state $|1'\rangle$ remains:
\begin{flalign}\label{eq:term0}
&\Omega_\textrm{MF}/N\rightarrow |t_0|\varphi^2-\theta\ln{\re^{-\beta\varepsilon_{1'}}}=|t_0|\varphi^2-\frac{\mu}{2}-\sqrt{\frac{\mu^2}{4}+t_0^2\varphi^2}\,.
\end{flalign}
Using expression (\ref{eq:simpl1}) we have:
\begin{flalign}\label{eq:term1}
&\Omega_\textrm{MF}/N=-\frac{(\mu+|t_0|)^2}{4|t_0|}\,.
\end{flalign}

At the same time, at $\varphi=0$, the ground state in the region $\mu'<0$ is the state $|0\rangle$ for $\mu<0$ and the state $|1\rangle$ for $\mu>0$. Then:
\begin{flalign}\label{eq:term2}
&\Omega_\textrm{MF}/N|_{\varphi=0}=
\left\{
  \begin{array}{ll}
    \phantom{-}0, & \hbox{$\mu < 0$,} \\
    -\mu, & \hbox{$\mu > 0$.}
  \end{array}
\right.
\end{flalign}

One can see that $\Omega_\textrm{MF}<\Omega_\textrm{MF}\big|_{\varphi=0}$; the SF phase is more stable in the interval $-|t_0|<\mu<|t_0|$. Derivatives
 ${\partial \Omega_\textrm{MF}}/{\partial \mu}$ and ${\partial \Omega_\textrm{MF}|_{\varphi=0}}/{\partial \mu}$ coincide in the limiting points $\mu=\pm|t_0|$. This verifies the second order of phase transitions to the phase with BE condensate here.

The function $\Omega_\textrm{MF}/N(\mu)$ has a similar character in the case $\mu'>U'$. Here:
\begin{flalign}\label{eq:term3}
&\Omega_\textrm{MF}/N=-\frac{(\mu-U'+|t_0|)^2}{4|t_0|} - \mu',\nonumber \\
&\Omega_\textrm{MF}/N\big|_{\varphi=0}=
\left\{
  \begin{array}{ll}
    -\mu', & \hbox{$\mu < U'$,} \\
    U'-\mu-\mu', & \hbox{$\mu > U'$.}
  \end{array}
\right.
\end{flalign}
Here, the second order phase transitions take place in the points $\mu=U'\pm|t_0|$.

The results of numerical calculations of function $\Omega_\textrm{MF}/N(\mu)$ in the case of intermediate values of $\mu'$ [performed with the help of the
earlier calculated $\varphi(\mu)$ dependences] are shown in figures~\ref{fig:new1}~(b)--\ref{fig:new5}~(b).
Here and hereafter, numerical values of parameters are given in the $U'$ units.


The special character of the dependences of the order parameter $\varphi$ on $\mu$ is quite understandable taking into account the change of the ground state
which occurs in the above mentioned intermediate region of $\mu'$ when $\varepsilon_{1'}=\varepsilon_{\tilde{1}'}$. The equation that follows from this condition
has a solution:
\begin{equation}
\varphi=\frac{\sqrt{\mu'(U'-\mu')(\mu-\mu')(\mu'-U'+\mu)}}{|t_{0}||2\mu'-U'|},
\label{eq:simpl3}
\end{equation}
which in the plane $(\mu,\varphi)$ describes the line that separates the areas with different ground states: the state $|\tilde{1}'\rangle $, ($|1'\rangle $)
to the left (right) of the curve (at the given value of $\mu'$).

At $\mu'=U'/2$, the line (\ref{eq:simpl3}) is vertical and passes through the point $\mu=U'/2$. At $\mu'<U'/2 $, it is bent to the left and  lies on the $x$-axis
when $\mu'\rightarrow 0 $; respectively, at $\mu'> U'/2 $ the bend is opposite, and curve lies on the $x$-axis at $\mu'\rightarrow U'$.

In the region where the ground state is $|1'\rangle$, the dependence $\varphi(\mu)$ is given by formula (\ref{eq:simpl1}) (see figure~\ref{fig:f1}), and when the
ground state is $|\tilde{1}'\rangle$, it is described by the formula (\ref{eq:simpl2}) (see figure~\ref{fig:f2}). Depending on the position of the line (\ref{eq:simpl3}),
certain fragments of graphs (\ref{eq:simpl1}) and (\ref{eq:simpl2}) remain on its both sides. This is illustrated in figures~\ref{fig:new3}~(a), \ref{fig:new4}~(a) and
\ref{fig:new5}~(a).
%

In the cases when the function $\varphi(\mu)$ has a reverse course, one can see the so-called ``fishtails'' in the behaviour of the grand canonical potential where the  intersection points of the lowest curves correspond to the first order phase transitions
(phase transitions on the other side or on both sides of the interval of non-zero $\varphi$ values are of the second order). The values of $\mu$ at which
the first order phase transitions exist, are shifted relatively to the spinodal line points (at $\mu'<U'/2$ to the left and at $\mu'>U'/2$ to the right).
As a result, the region of SF phase existence at $T=0$ is wider than the region limited by spinodals.

\section{Phase diagrams at $T=0$}
\label{sec:7}

\begin{figure}[!t]
\centerline{
\includegraphics[width=0.56\textwidth]{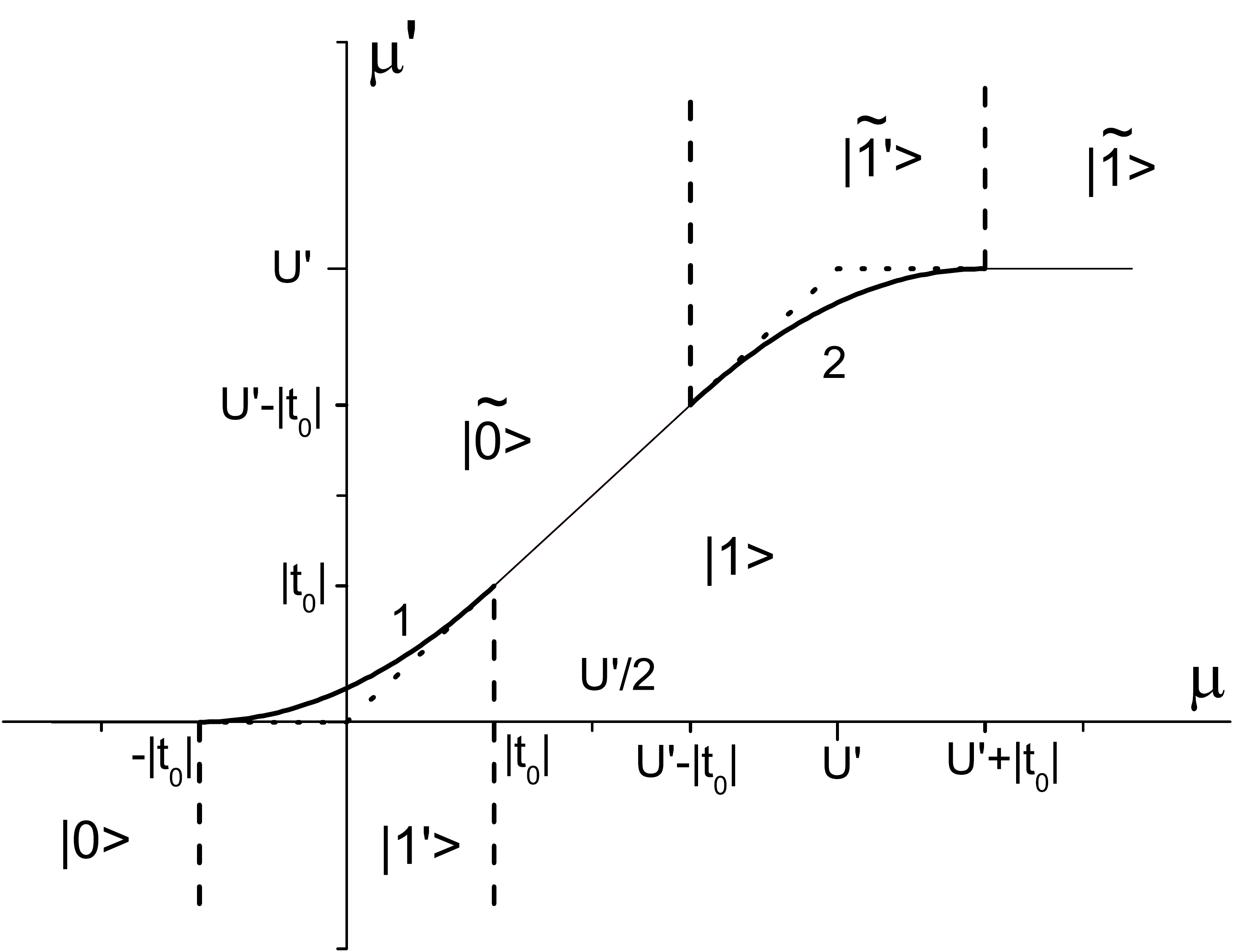}
}
\caption{$(\mu, \mu')$ diagram at $|t_0|<U'/2$\protect\footnotemark.}
\label{fig:mumup1}
\end{figure}

A more detailed analysis shows that at $|t_{0}|<U'/2$ in the $0<\mu'<|t_{0}|$ region, the phase transition of the first order to the SF phase is determined by the intersection of the branches of the grand canonical potential
\begin{equation}
\frac{\Omega_\textrm{MF}}{N}{\Big|}_{|\widetilde{0}\rangle}=-\mu'
\label{ista8.10}
\end{equation}
and
\begin{equation}
\frac{\Omega_\textrm{MF}}{N}{\Big|}_{|1'\rangle}=-\frac{(\mu+|t_{0}|)^{2}}{4|t_{0}|}
\label{ista8.11}
\end{equation}
and is associated with the change ($|\widetilde{0}\rangle\rightarrow |1'\rangle$) of the ground state of boson-fermion system. Consequently, the relation
\begin{equation}
\mu'=\frac{(\mu+|t_{0}|)^{2}}{4|t_{0}|},
\label{ista8.12}
\end{equation}
arises which is an equation of the phase transition line on the plane $(\mu,\mu')$. In figure~\ref{fig:mumup1}, where the regions of the existence of different
ground states are shown in $(\mu,\mu')$ diagram, this curve is depicted by heavy solid line. SF phase exists here in the regions marked as $|1'\rangle$ and
$|\widetilde{1'}\rangle$; to distinguish between these two cases, we shall use, respectively, the notations SF$^{|1'\rangle}$ and SF$^{|\widetilde{1'}\rangle}$.

\begin{figure}[!t]
\centerline{
\includegraphics[width=0.56\textwidth]{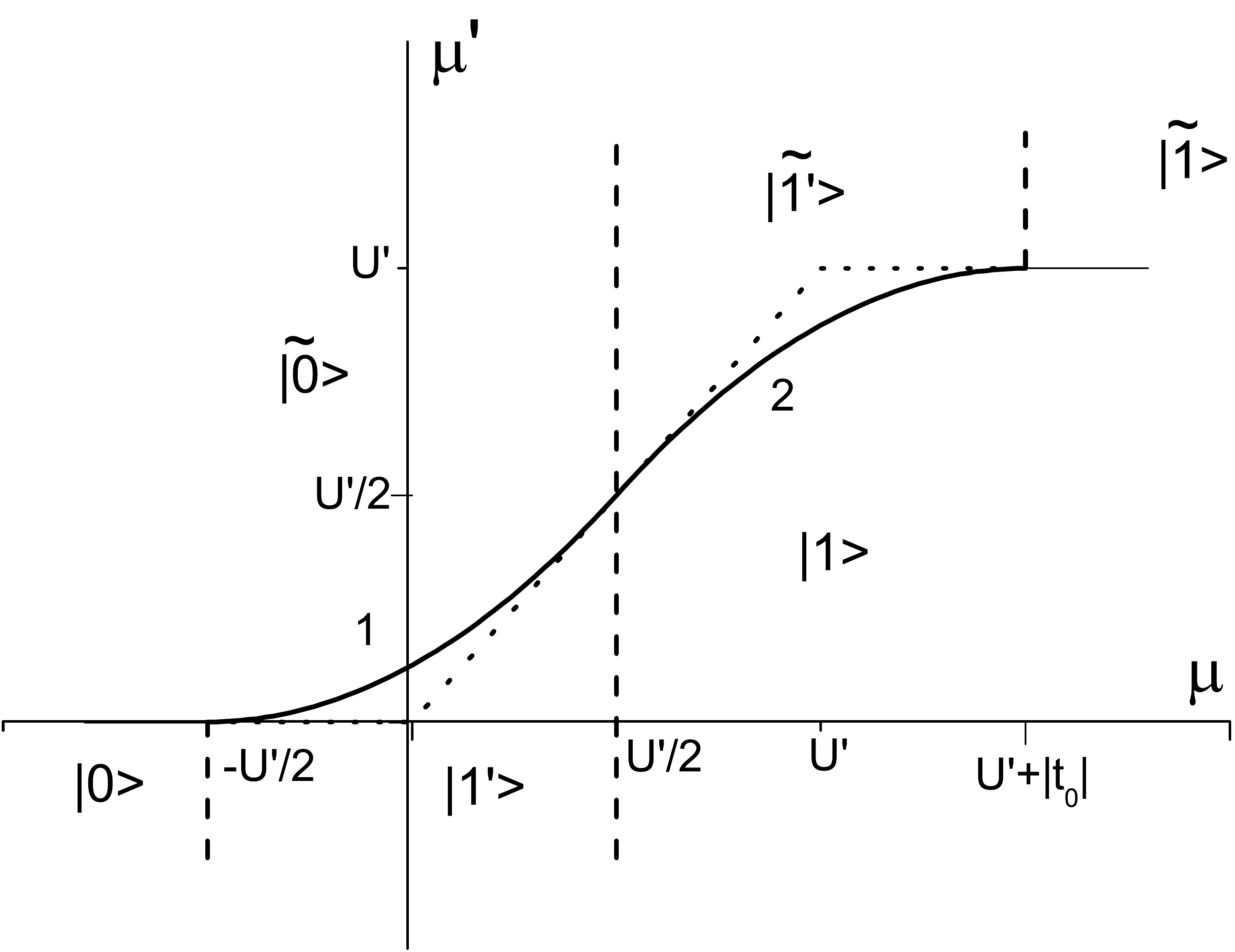}
}
\caption{$(\mu, \mu')$ diagram at $|t_0|=U'/2$.}
\label{fig:mumup2}
\end{figure}

In the case when $U'/2<|t_{0}|<U'$, the curve (\ref{ista8.12}) proceeds to describe the transition of the 1st order until $0<\mu'<(U')^{2}/4|t_{0}|$.
At the same time, it remains to the left of spinodal line (figure~\ref{fig:mumup2}). When $(U')^{2}/4|t_{0}|<\mu'<U'-(U')^{2}/4|t_{0}|$, the line of the first order
transitions is placed between spinodals. The latter are described by equations $\mu=U'-|t_{0}|$ and $\mu=|t_{0}|$; at such values of $\mu$, the second order phase transitions to the SF phase occur. For $U'-|t_{0}|<\mu<\mu'$, this is a phase
SF$^{|\widetilde{1'}\rangle}$ and for $\mu'<\mu<|t_{0}|$~--- a phase SF$^{|1'\rangle}$ (superscripts indicate the states of the system at $T=0$).
The transition between these two phases, which is of the first order, is defined by the equality of grand canonical potentials
${\Omega_\textrm{MF}}/{N}{\big|}_{|1'\rangle}$ and ${\Omega_\textrm{MF}}/{N}{\big|}_{|\tilde{1}'\rangle}$, here,
\begin{equation}
\frac{\Omega_\textrm{MF}}{N}{\Big|}_{|\tilde{1}'\rangle}=
-\frac{(\mu-U'+|t_{0}|)^{2}}{4|t_{0}|}-\mu'.
\label{ista8.13}
\end{equation}
The line that describes this transition on the plane $(\mu,\mu')$ is:
\begin{equation}
\mu'=\frac{U'}{4|t_{0}|}(2\mu+2|t_{0}|-U').
\label{ista8.14}
\end{equation}
In the region $U'-(U')^{2}/4|t_{0}|<\mu'<U'$, the line of the first order phase transition continues; it separates the phase SF$^{|\widetilde{1'}\rangle}$
from the normal phase (for the latter, the ground state is $|1\rangle$). In this case, the equation of the phase equilibrium curve follows from the condition:
\begin{equation}
\frac{\Omega_\textrm{MF}}{N}{\Big|}_{\tilde{1}'\rangle}
=\frac{\Omega_\textrm{MF}}{N}{\Big|}_{|1\rangle}\,,
\label{ista8.15}
\end{equation}
where ${\Omega_\textrm{MF}}/{N}{\big|}_{|1\rangle}=-\mu$ and has the form
\begin{equation}
\mu'=\mu-\frac{(\mu-U'+|t_{0}|)^{2}}{4|t_{0}|}\,.
\label{ista8.16}
\end{equation}
On the $(\mu,\mu')$ plane, this line is positioned to the right of the spinodal line.

\begin{figure}[!t]
\centerline{
\includegraphics[width=0.6\textwidth]{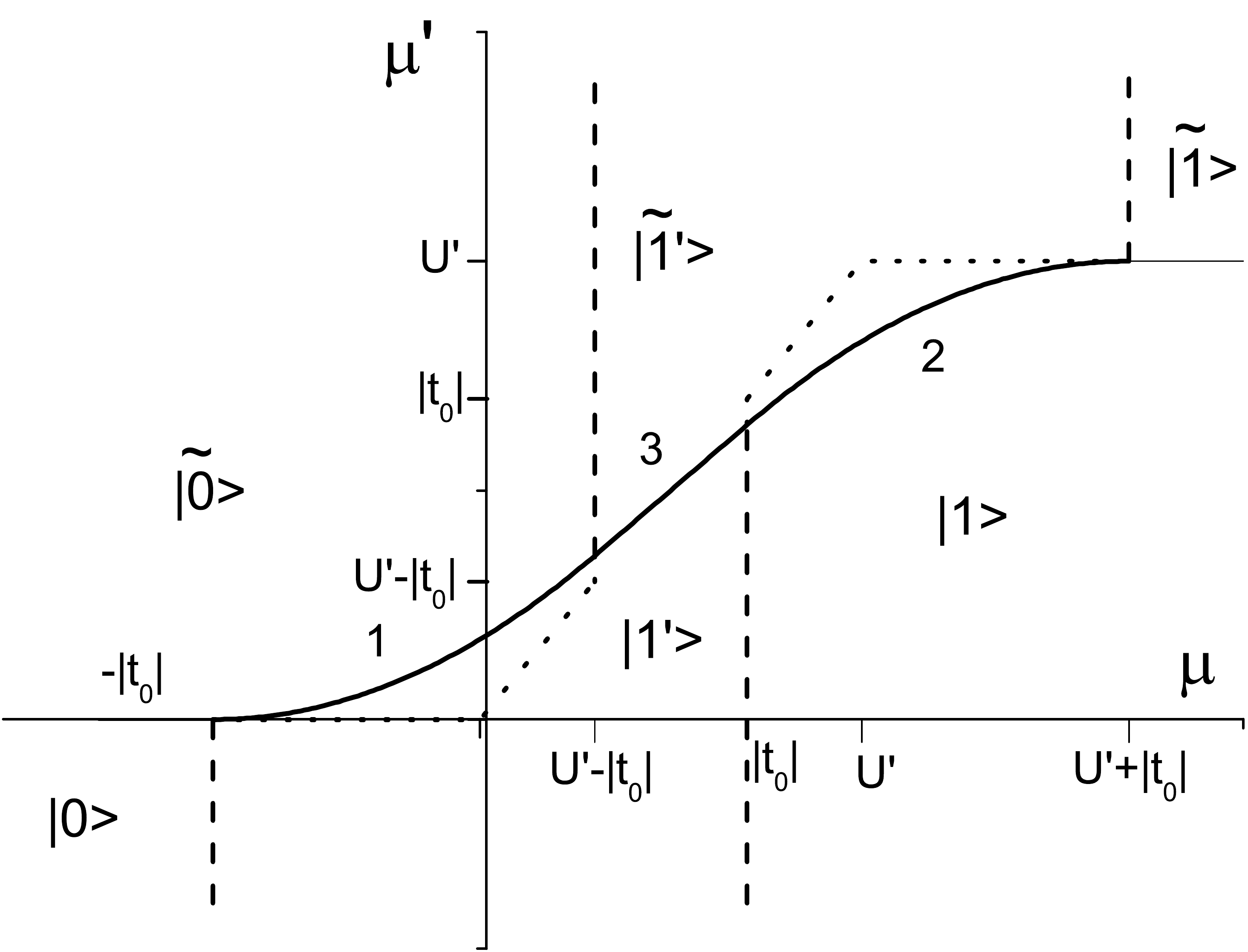}
}
\caption{$(\mu, \mu')$ diagram at $U'/2<|t_0|<U'$.}
\label{fig:mumup3}
\end{figure}

\begin{figure}[!t]
\centerline{
\includegraphics[width=0.6\textwidth]{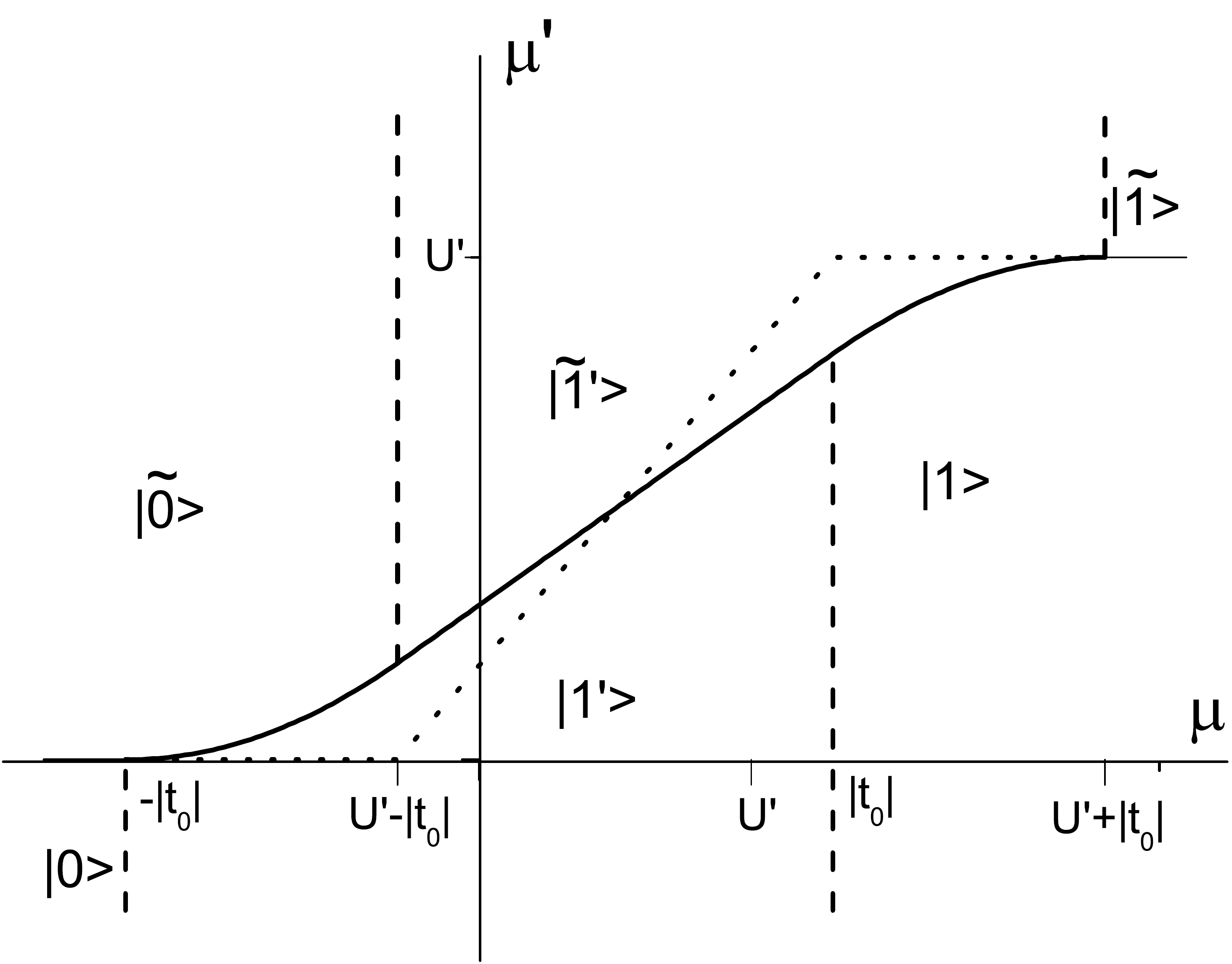}
}
\caption{$(\mu, \mu')$ diagram at $|t_0|>U'$.}
\label{fig:mumup4}
\end{figure}

\begin{figure}[!t]
\centerline{
\includegraphics[width=0.45\textwidth]{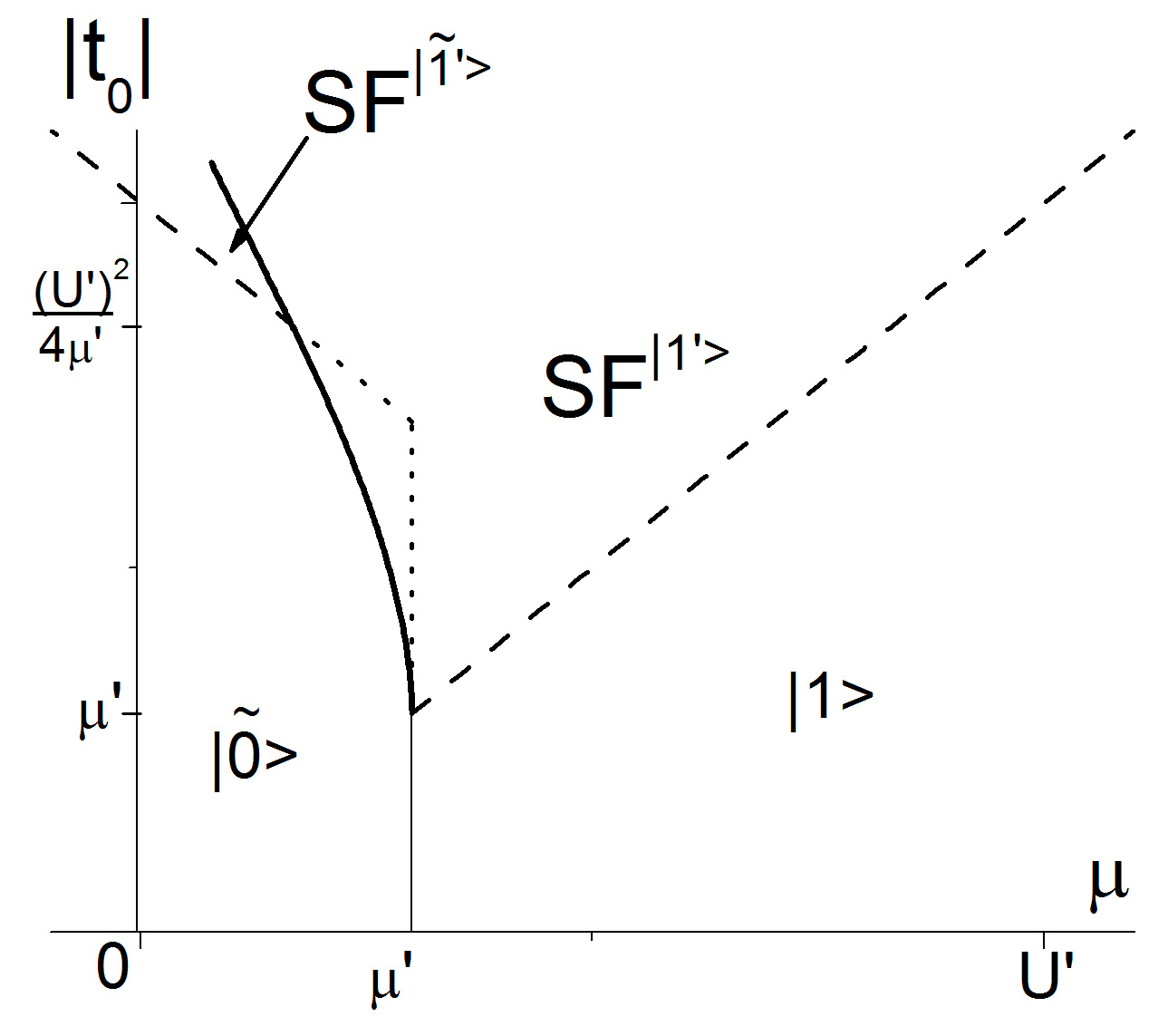}
\hspace{5mm}
\includegraphics[width=0.47\textwidth]{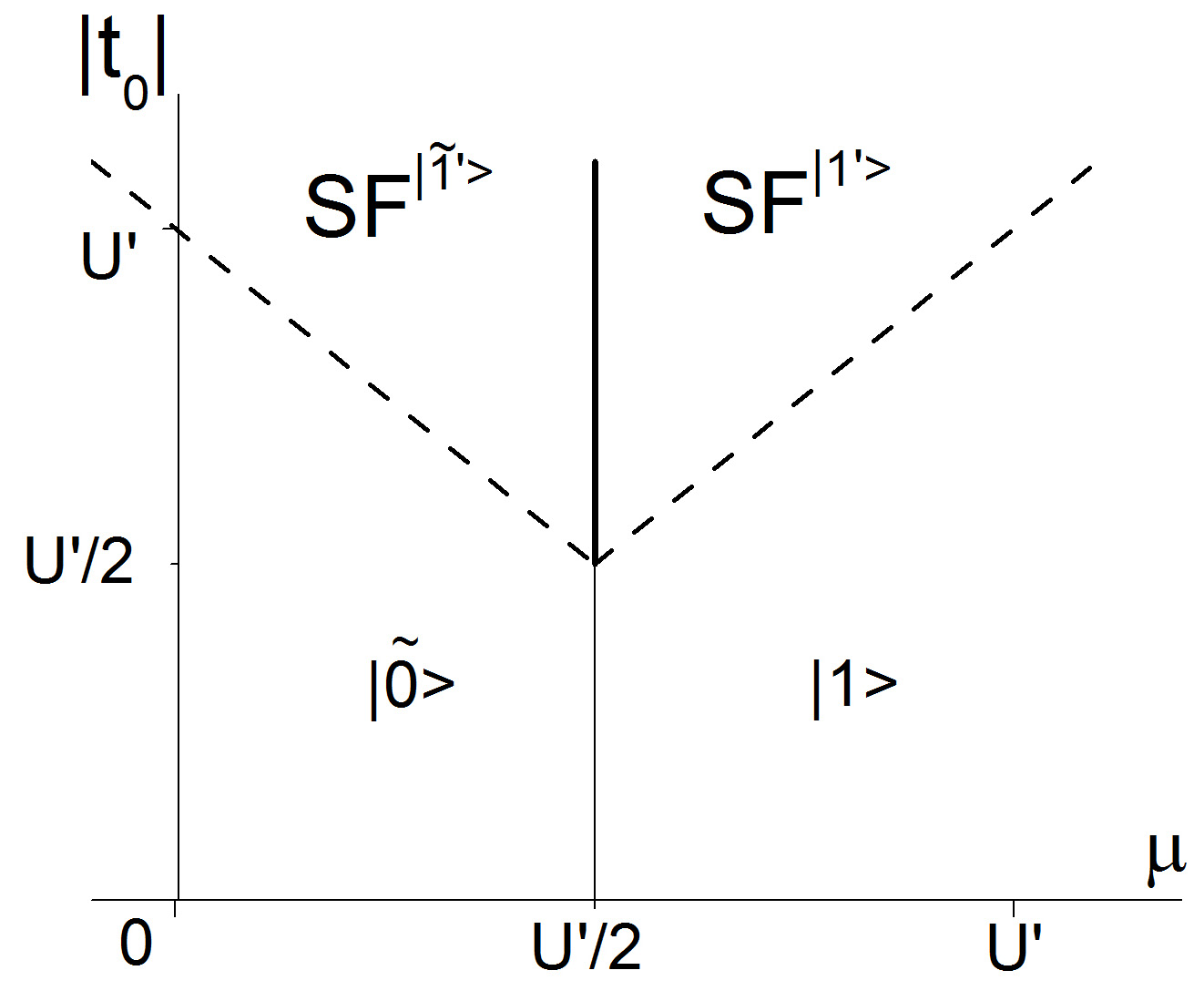}
}
\parbox[t]{0.49\textwidth}{%
\caption{$(\mu, |t_0|)$ diagram at $0<\mu'<U'/2$.}
\label{fig:tomu1}
}%
\hfill%
\parbox[t]{0.49\textwidth}{%
\caption{$(\mu,|t_0|)$ diagram at $\mu'=U'/2$.}
\label{fig:tomu2}
}
\end{figure}

\begin{figure}[!t]
\centerline{\includegraphics[width=0.52\textwidth]{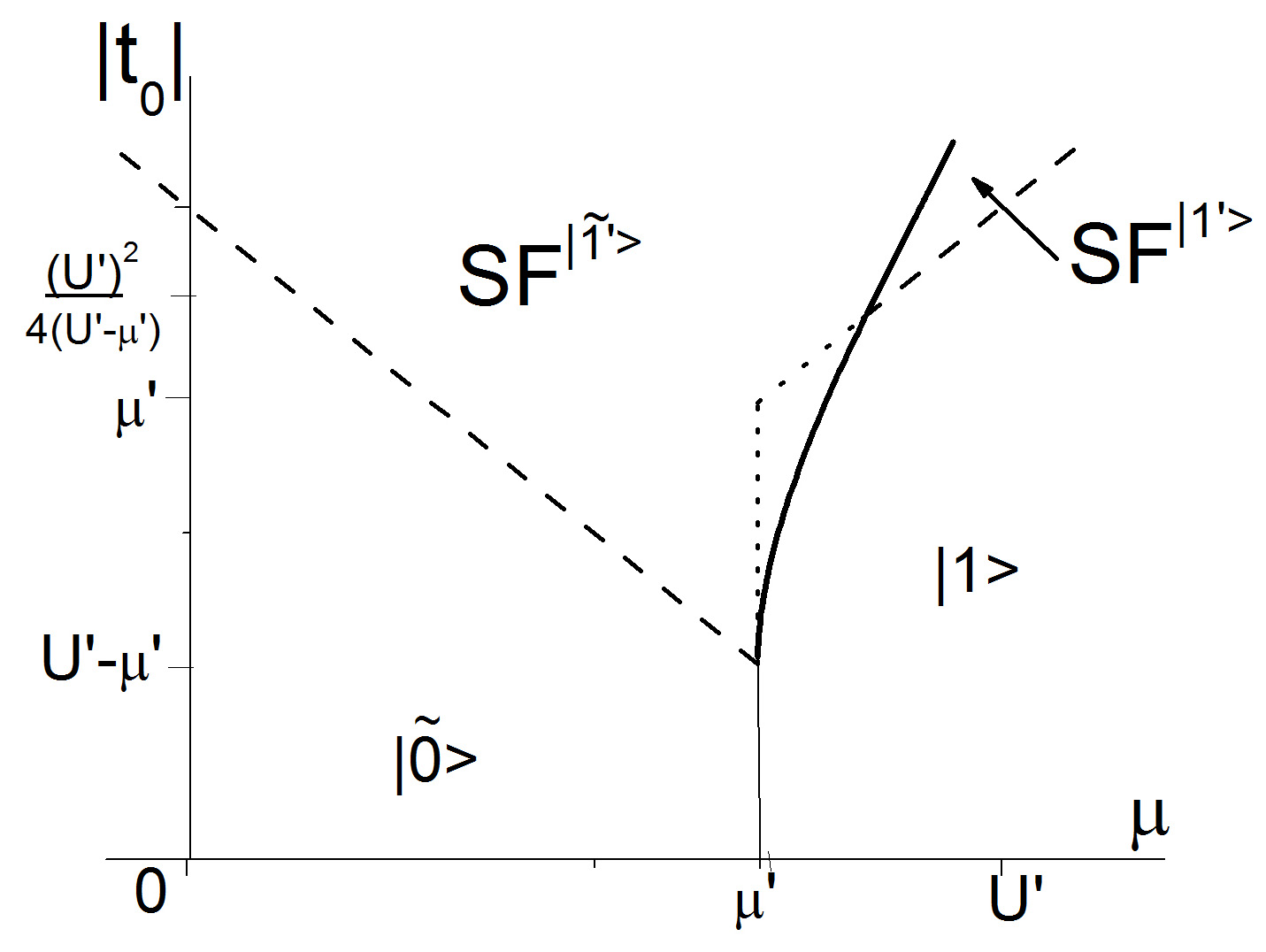}}
\caption{$(\mu,|t_0|)$ diagram at $U'/2<\mu'<U'$.}
\label{fig:tomu3}
\end{figure}

\footnotetext{Here, as well as in figures \ref{fig:mumup2}--\ref{fig:tomu3}, the following notations are used: heavy solid line for the 1st order PT; dashed line for the 2nd order PT;
dotted line for the spinodal; fine solid line for the border between different ground states which belongs to normal phase.}

Equations (\ref{ista8.12}), (\ref{ista8.14}) and (\ref{ista8.16}) define the lines of phase transitions of the 1st order at $|t_{0}|>U'$ as well. A full set of phase diagram $(\mu,\mu')$ types is supplemented by diagrams shown in figures~\ref{fig:mumup3} and \ref{fig:mumup4}. It should be pointed out that
the continuous line of PT of 1st order in the whole region $ 0<\mu'<U'$ separates the states (phases) with one fermion (``tilded'' ones) and the states without
fermions (``untilded'' ones). The distinction between the superfluid phases SF$^{|\widetilde{1'}\rangle}$ and SF$^{|1'\rangle}$ is related to this fact. In the first case, BE
condensate exists at the full fermion filling $\left(\bar{n}_{f}=1\right)$, and in the second one, at the absence of fermions $(\bar{n}_{f}=0)$.

At nonzero temperatures, the dependence of $\bar{n}_{f}$ on $\mu'$ at crossing the line of PT of 1st order is smooth. The concentration of
fermions $\bar{n}_{f}$  gradually reduces in ``tilded'' areas and correspondingly increases in a similar way in ``untilded'' areas. The principal difference
between the phases of SF$^{|\widetilde{1'}\rangle}$ and  SF$^{|1'\rangle}$ vanishes.
%

The above described effect of the phase transition order change also takes place when chemical potential $\mu'$ is positioned in the middle of the $[0, U']$ interval (see figure \ref{fig:mumup3}). The point $\mu'=U'/2$ is a special one. With a decrease of $\mu'$, the fragmentation of SF region into two parts
takes place at this point.

%
%
Having  the phase diagrams $(\mu,\mu')$ built for various values of $|t_{0}|$, one can pass to diagrams on the plane $(\mu,|t_0|)$.
Using formulae (\ref{ista8.12}), (\ref{ista8.14}) and (\ref{ista8.16}) it is easy to get the relations between $\mu$ and $|t_{0}|$
(for fixed values of $\mu'$) on the lines of phase transitions of the first order:
\begin{equation}
\mu=\sqrt{4|t_{0}|\mu'}-|t_{0}|,
\label{ista8.17}
\end{equation}
--- when $\mu'<U'/2$;
\begin{equation}
\mu=U'+|t_{0}|-\sqrt{4|t_{0}|(U'-\mu')},
\label{ista8.18}
\end{equation}
--- when $\mu'>U'/2$; and
\begin{equation}
\mu=\frac{U'}{2}-|t_{0}|+\frac{2|t_{0}|\mu'}{U'}
\label{ista8.19}
\end{equation}
--- in both these cases when the phase transition occurs within SF-phase region.

Using formulae (\ref{ista8.17}), (\ref{ista8.18}) and (\ref{ista8.19}), and  the equation for spinodals at $T=0$, we can get the diagrams shown
in figures~\ref{fig:tomu1}, \ref{fig:tomu2} and \ref{fig:tomu3}. In contrast to the $(\mu,|t_{0}|)$ diagram for pure bosonic case, figure~\ref{fig:tomu2}, known
for a hard-core boson system (see, for example, \cite{micnas,ista2}), these diagrams are asymmetric. At the smaller (or larger) values of $\mu$, the
phase transition  (under the effect of fermions) changes its order from the second to the first one, depending on the value of the chemical potential $\mu'$.

There also appears the minimum value for bosons transfer parameter ($|t_{0}|_\textrm{min}=\mu'$
at $\mu'<U'/2$ or $|t_{0}|_\textrm{min}=U'-\mu' $ at $\mu'>U'/2$). SF phase can exist only if $|t_{0}|>|t_{0}|_\textrm{min}$.

\section{Phase separation at a fixed boson chemical potential}
\label{sec:8}

\begin{figure}[!b]
\centerline{\includegraphics[width=0.55\textwidth]{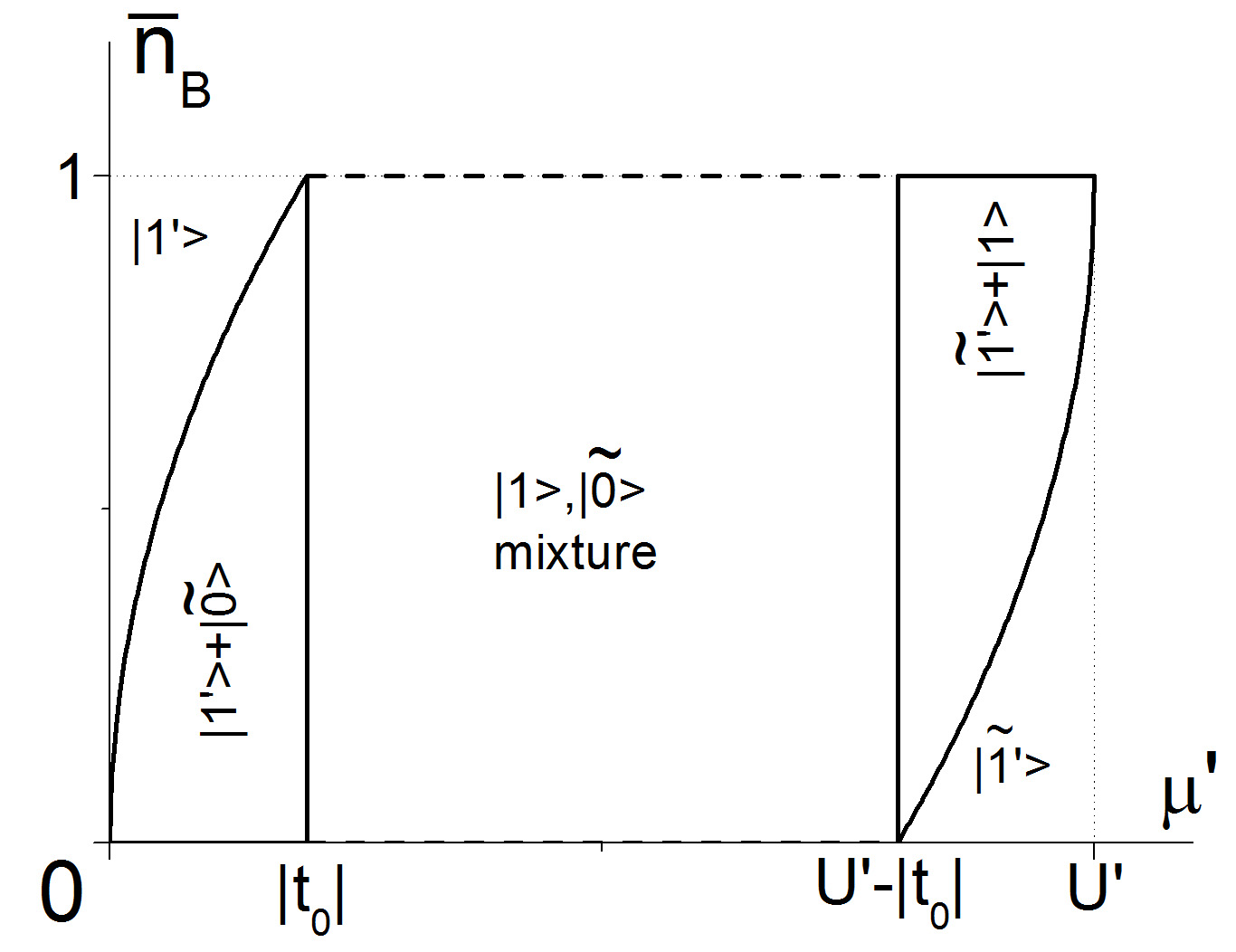}}
\caption{$(\mu',\bar{n}_\textrm{B})$ diagram at $|t_0|<U'/2$.}
\label{fig:nbmu1}
\end{figure}

\begin{figure}[!t]
\centerline{\includegraphics[width=0.55\textwidth]{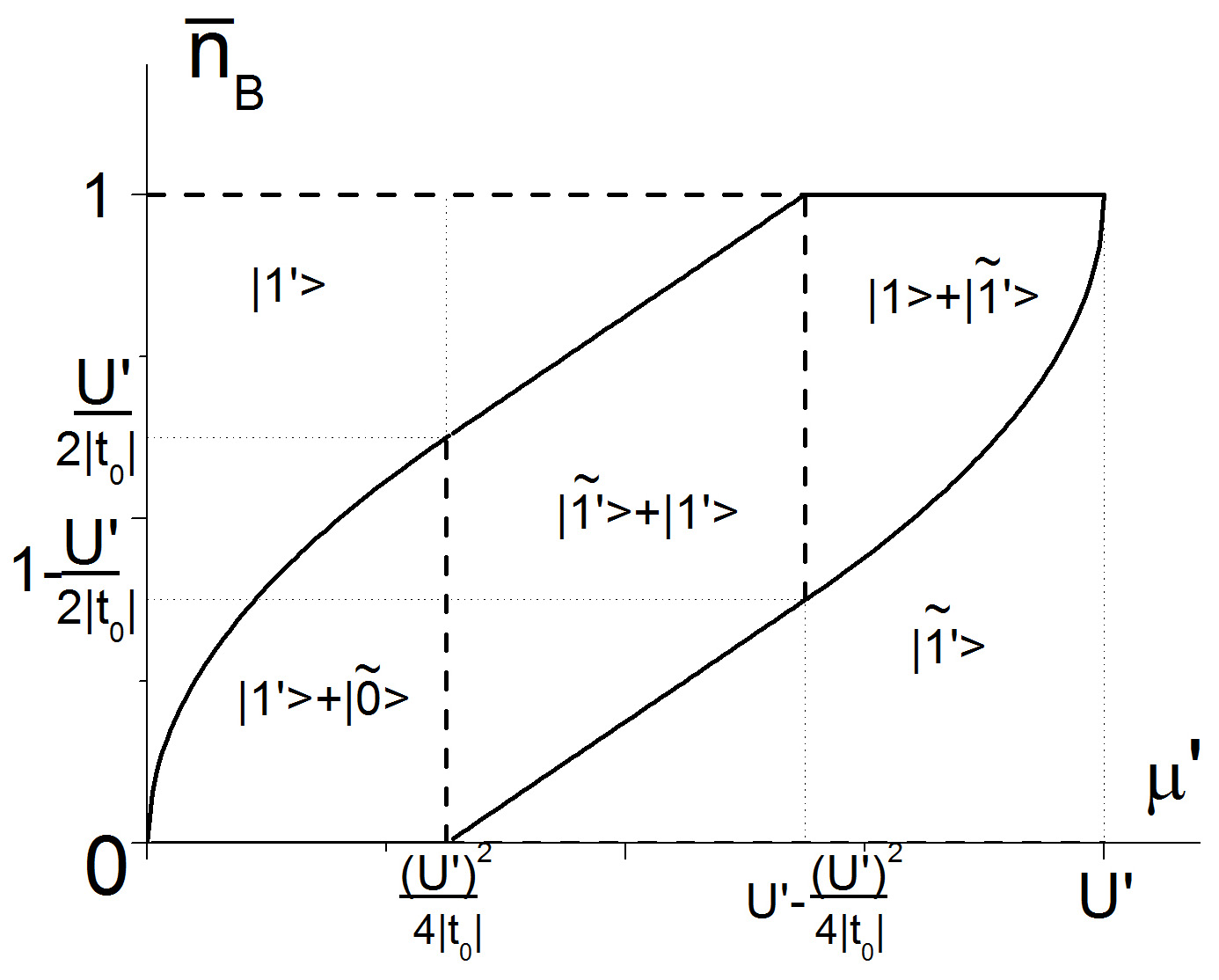}}
\caption{$(\mu',\bar{n}_\textrm{B})$ diagram at $U'/2<|t_0|<U'$.}
\label{fig:nbmu2}
\end{figure}

Besides a jump of a number of fermions, there is also a jump-like behaviour of boson concentration on the line of phase transition of the 1st order,
 $\bar{n}_\textrm{B}=-\partial\left(\Omega_\textrm{MF}/N\right)/\partial\mu$. At zero temperature
\begin{align}
&\bar{n}_\textrm{B}{\big|}_{|\tilde{0}\rangle}=0,& & \bar{n}_\textrm{B}{\big|}_{|1\rangle}=1, \nonumber\\
&\bar{n}_\textrm{B}{\big|}_{|\tilde{1}\rangle}=1, & & \bar{n}_\textrm{B}{\big|}_{|{0}\rangle}=0, \nonumber\\
&\bar{n}_\textrm{B}{\big|}_{|\tilde{1}'\rangle}=\frac{\mu-U'+|t_{0}|}{2|t_{0}|}, \hspace{-2cm}& &
\bar{n}_\textrm{B}{\big|}_{|{1'}\rangle}=\frac{\mu+|t_{0}|}{2|t_{0}|}.
\label{ista9.1}
\end{align}
Using these relations,  the limiting values of $\bar{n}_\textrm{B}$ (from the left or right sides of the transition lines),
can be written for a given $\mu'$. Using the formulae (\ref{ista8.12}), (\ref{ista8.14}) and (\ref{ista8.16}), we find:
\begin{enumerate}
\item[1)] on the line (\ref{ista8.12})
\begin{equation}
\bar{n}_\textrm{B}{\big|}_{l}=0,\qquad \bar{n}_\textrm{B}{\big|}_{r}=\sqrt{\mu'/|t_{0}|};
\label{ista9.2}
\end{equation}
\item[2)] on the line (\ref{ista8.14})
\begin{equation}
\bar{n}_\textrm{B}{\big|}_{l}=\frac{\mu'}{U'}-\frac{U'}{4|t_{0}|},\qquad \bar{n}_\textrm{B}{\big|}_{r}=\frac{\mu'}{U'}+\frac{U'}{4|t_{0}|};
\label{ista9.3}
\end{equation}
\item[3)] on the line (\ref{ista8.16})
\begin{equation}
\bar{n}_\textrm{B}{\big|}_{l}=1-\sqrt{\frac{U'-\mu'}{|t_{0}|}},\qquad \bar{n}_\textrm{B}{\big|}_{r}=1.
\label{ista9.4}
\end{equation}
\end{enumerate}
The dependences $\bar{n}_\textrm{B}{\big|}_{l,r}$ on $\mu'$ are presented graphically in figures~\ref{fig:nbmu1} and \ref{fig:nbmu2}; instances $|t_{0}|<U'/2 $ and $U'/2<|t_{0}|<U'$ are shown separately.
In thermodynamical regime of given values of chemical potentials $\mu$ and $\mu'$, these plots illustrate the jump of the concentration of bosons in different points
of the phase equilibrium curve. On the other hand, in the regime of fixed $\bar{n}_\textrm{B} $ and $\mu'$, they can be interpreted as diagrams describing the phase
separation on the regions with different concentrations $\bar{n}_\textrm{B}$.

The values $\bar{n}_\textrm{B}{\big|}_{l}$ and $\bar{n}_\textrm{B}{\big|}_{r}$ correspond in this case to different phases (states), into which the system segregates, depending on the value of chemical potential $\mu'$. This is shown in diagrams~\ref{fig:nbmu1} and \ref{fig:nbmu2} (the separation regions are enclosed by solid lines); the phases into which the separation takes place are also shown.

In the case $U'/2<|t_{0}|<U'$ (also at $|t_{0}|>U'$), the separation occurs throughout the whole interval of $0<\mu'<U'$. When $|t_{0}|<U '/ 2 $,
the system decomposes into separate phases only at the $\mu'$ values in regions $\mu'<|t_0|$ and $U'-|t_0|<\mu'<U'$. There is no separation
 in the central region $|t_{0}|<\mu'<U'-|t_{0}|$ in this case; here, at a given fractional value of $\bar{n}_\textrm{B}$, the system exists in a mixed state, where the lattice
 sites are in the state $|\tilde{1}\rangle$ (boson and fermion on site) or $|0\rangle$ (unfilled site) with probabilities determined by their weights
 ($\bar{n}_\textrm{B}$ or $1-\bar{n}_\textrm{B}$, respectively).

\section{Conclusions}
\label{sec:9}

To describe phase transitions in the boson-fermion mixtures in optical lattices
we used the Bose-Fermi-Hubbard model in the mean-field and hard-core boson approximations, for the case of infinitely
small fermion transfer and repulsive on-site boson-fermion interaction.
Our aim was to study the conditions, at which the MI-SF transition in such a model occurs, in the case when the fermion hopping between lattice sites can be
neglected. The approach used in this work does not use the traditional scheme of mean-field approximation based on the decoupling of the on-site interaction
$U'n_i^\textrm{b}n_i^\textrm{f}$. Instead, the formalism of Hubbard operator acting on the $|n_i^\textrm{b},n_i^\textrm{f}\rangle$ basis of states is employed; this makes it possible
to exactly take into account the boson-fermion interaction $U'$ (the case of repulsive interaction $(U'>0)$ is considered in this work).
The single-site problem is formulated with only one self-consistency parameter $\varphi$ ($\varphi=\langle b_i\rangle=\langle b^+_i\rangle$),
and the mean-field approach is used to describe the BE condensation.

On-site boson interaction $U$ is treated as repulsive and infinitely large ($U>0$,  $U\rightarrow\infty$); this imposes a restriction on the occupation numbers
of bosons ($n_i^\textrm{b}=0$ or 1). Nevertheless, this approximation makes it possible, as is known for the pure Bose-Hubbard model, to describe the MI-SF transition
in close vicinity of the $\mu=nU$ points (where $n$ are integer numbers) in the case of finite values of $U$. The investigation is performed in a thermodynamical regime of fixed values of chemical potentials of bosons ($\mu$) and fermions~($\mu'$).

The equilibrium values of the order parameter $\varphi$ (related to the appearance of SF phase) were found from the global minimum
condition of grand canonical potential $\Omega$ and, in parallel, the averages of creation and destruction operators of bosons $\langle b\rangle$ and
$\langle b^+\rangle$ were calculated. From the obtained equation, using substitution $\varphi\rightarrow 0$, we get the condition of 2nd order phase transition to SF phase (if this transition is possible).  In general, it is the condition of instability of normal (MI) phase with respect to the Bose-Einstein condensate appearance. This equation is the same as as the one obtained earlier from the condition of divergence of the bosonic Green's function (calculated in the random phase approximation) at $\omega=0$, ${\bf q}=0$ (see \cite{kras}).

The spinodal lines are calculated  at $T=0$, and corresponding phase diagrams on the ($\mu,\mu'$) planes are built. Analyzing their shape, different cases
are separated depending on the value of the chemical potential of fermions $\mu'$, and thus the corresponding phase diagrams are built. Moreover, the dependences of the
order parameter $\varphi$ and the grand canonical potential $\Omega$ on $\mu$ (at different chemical potential $\mu'$ values) are derived.

Considering the order parameter dependence upon chemical potential of bosons $(\mu)$ we found that in the region of intermediate values of $\mu'$ (especially at
$0\leqslant \mu'\leqslant U'$) the competition between ``tilded'' and ``untilded'' states leads to a deformation of the curve $\varphi(\mu)$. These cases are distinguished
when such a dependence has a reverse course and S-like behaviour. This is an evidence of the possibility of the first order phase transition between MI and SF phases
(rather than the second order transition). This conclusion was confirmed by calculation of grand canonical potential $\Omega_\textrm{MF}(\mu)$ as function of $\mu$.
As a result, the region of the existence of SF phase at $T=0$ is wider than the limited region by spinodals. The above described effect of the phase transition order
change disappears when chemical potential $\mu'$ takes the values $\mu'<0$ and $\mu'>U'$. In the first case, there are no fermions  $(\overline{n}_\textrm{F}=0)$
while in the second case the fermion states are fully occupied $(\overline{n}_\textrm{F}=1)$. At $\overline{n}_\textrm{F}=0$, the model is reduced to the pure hard-core boson model with
the phase transitions of the 2nd order; at $\overline{n}_\textrm{F}=1$, the picture of a MI-SF transition is the same but the chemical potential of bosons
is shifted $(\mu\rightarrow \mu+U')$.

The $[0, U']$ interval of the $\mu'$ values corresponds to the fractional $(0<\overline{n}_\textrm{F}<1)$ fermion concentration.
BE condensation taking place in this case is influenced  by the states which have different number of fermions. The point $\mu'=U'/2$ is a special one.
With a decrease of $\mu'$, the fragmentation of SF region into two parts takes place at this point.

The obtained results are illustrated with the help of $(\mu,\mu')$ and $(\mu,|t_0|)$ phase diagrams. It should be emphasized that in the case $|t_0|>U'/2$, when ${(U')^2}/({4|t_0|})<\mu'<U'-{(U')^2}/({4|t_0|})$, the
1st order phase transition line is placed inside the SF region. It divides this region into parts with BE condensate of different type (SF$^{|\widetilde{1'}\rangle}$ region
when all fermion states are filled, and SF$^{|1'\rangle}$ region when there are no fermions). The asymmetry of $(\mu, |t_0|)$ phase diagrams (contrary to the pure
hard-core boson system) is another interesting feature. There also exists a minimum value of the transfer parameter $|t_0|$ necessary for the SF phase existence.

We have also briefly considered the regime of a fixed boson concentration. In this case, the phase separation effect is possible. The phase diagrams $(\mu', \overline{n}_\textrm{B})$
are obtained where the separation regions are shown. Depending on $\mu'$ value, the system can segregate into the regions with different $\overline{n}_\textrm{B}$ concentrations
and different phases (MI and SF or SF$^{|\widetilde{1'}\rangle}$ and SF$^{|1'\rangle}$).

It is not easy to directly compare the obtained phase diagrams with the available data concerning the thermodynamics of the BFH model. In most cases, the investigations
were performed for another thermodynamical regime, namely the regime of fixed fermion concentration (besides the given chemical potential of bosons). Starting
with our scheme, the transition to the regime of the fixed $\overline{n}_\textrm{F}$ values could be made with the help of Legendre transformation
$\Omega/N\rightarrow \widetilde{\Omega}/N=\Omega/N+\mu'\overline{n}_\textrm{F}$ and subsequent transition to the new thermodynamical variables. Nevertheless, one can see that the states
with a certain fractional value of $\overline{n}_\textrm{F}$ are positioned on the $(\mu, \mu')$ diagram at $T=0$ on the curve that separates the ``tilded'' and ``untilded'' regions
(the jump-like change from $\overline{n}_\textrm{F}=1$ to $\overline{n}_\textrm{F}=0$ value takes place at the crossing over this line). Moving along the line (at the change of $\mu$) we shall
pass through the intervals of $\mu$ values corresponding to the regions of the SF phase existence. When $T=0$ and $\overline{n}_\textrm{F}>1/2$ ($\overline{n}_\textrm{F}<1/2$), the above mentioned
curve will be placed a little higher (lower) of the zero-temperature curve. The symmetry of the diagram will be broken, and this leads to a situation when at a
decrease of $|t_0|$, the SF phase disappears in the region near the $\mu=0$ point sooner (or later) than near the $\mu=U'$ point. Although this conclusion is qualitative, it can be considered
as a confirmation and explanation of the results obtained in \cite{kras} for the full BFHM in the $t_\textrm{F}=0$ case at finite temperatures in the regime $\overline{n}_\textrm{F}=\textrm{const}$.
We should stress that in \cite{kras}, as in a lot of works in this field (see for example \cite{fehr1,polak}, the phase diagrams were built basing on the conditions of the SF phase instability (the latter
is determined by spinodals).

The fact of the minimum $|t_0|$ value existence necessary for the appearance of the SF phase at the presence of fermions (which is seen on $(\mu, |t_0|)$ diagrams,
figures~\ref{fig:tomu1}--\ref{fig:tomu3}) is a consequence of an exact treatment of the boson-fermion interaction $U'$ in our approach.
However, when such an interaction is taken into account basing on a simple linearization scheme (in the spirit of Hartree-Fock mean-field decoupling),
the minimum value $|t_0|_\textrm{min}$ is equal at $T=0$ to zero \cite{fehr1,fehr2,polak,refa}. This value is reached at certain $\mu=\mu^*$ and the effect
of fermions consists in the shift of the $\mu^*$ point \cite{polak}. We have a similar effect  in our case: the position of non-zero $|t_0|_\textrm{min}$ as function
of $\mu$ depends on $\mu'$ (see figures~\ref{fig:tomu1}, \ref{fig:tomu3}). In this regard, the investigations of changes that could appear at finite temperatures and the comparison of different approaches  turn out to be an interesting task (at $T\neq 0$ $|t_0|_\textrm{min}$ is non-zero even in a simple mean-field approximation). For a heavy-fermion limit, as was shown in \cite{refa},
the SF phase region becomes broader near $|t_0|_\textrm{min}$ point and the $|t_0|_\textrm{min}$ value increases at the temperature growth.

It is also interesting to study, in the fixed chemical potentials ($\mu$ and $\mu'$) regime, the phase transition picture at the finite fermion hopping. Such an
investigation was done in many works for a given fermion concentration; a rather full analysis was performed in \cite{buk} using a simple version of the mean-field
approach for $U_\textrm{bf}$ interaction. The possibility of the band fermion pairing leads at certain conditions to the appearance of phases with the pair fermion condensate.
It is necessary, however, to exceed some critical value of the hopping parameter $t_\textrm{F}$ for their thermodynamical preference \cite{buk,anders}.

Summarizing, we can make the following main conclusion. The phase transition to the SF-phase in a mixed system (consisting of bosons and heavy fermions), described by the hard-core model
and considered in the $\mu=\textrm{const}$ and $\mu'=\textrm{const}$ regime, becomes of the 1st order in  such a region of the chemical potential values where the BE condensation is
influenced by competition between filled and empty fermion states. As a consequence, the phase separation at fixed concentrations of particles can appear.

\ukrainianpart

\title{Фазові переходи у моделі Бозе-Фермі-Хаббарда в наближенні важких ферміонів: підхід жорстких бозонів}
\author{І.В. Стасюк, В.О. Краснов}
\address{
Інститут фізики конденсованих систем НАН України, вул.~І.~Свєнціцького, 1,
79011 Львів, Україна
}

\makeukrtitle

\begin{abstract}
\tolerance=3000%
Досліджено фазові переходи в моделі Бозе-Фермі-Хаббарда в наближенні середнього поля та жорстких бозонів при врахуванні  одновузлової бозон-ферміонної взаємодії типу відштовхування та у випадку безмежно малого переносу ферміонів. Проаналізовано поведінку параметра порядку бозе-конденсату
та термодинамічного потенціалу як функцій хімічного потенціалу бозонів при нульовій температурі. Встановлено можливість зміни роду фазового переходу до
 надплинної фази у режимі заданих значень хімічних потенціалів бозе- та фермі-частинок. Побудовано відповідні фазові діаграми.
\keywords модель Бозе-Фермі-Хаббарда, жорсткі бозони, бозе-конденсат, фазові переходи, фазові діаграми

\end{abstract}

\end{document}